\documentclass[aps,prd,superscriptaddress,nofootinbib,11pt]{revtex4}
\usepackage[english]{babel}
\usepackage[utf8]{inputenc}
\usepackage{graphicx}   
\usepackage{slashed}
\usepackage{epstopdf}
\usepackage{verbatim}   
\usepackage{color}      
\usepackage{subfigure}  
\usepackage{multirow}
\usepackage{hyperref}   
\usepackage{float}
\usepackage{epsfig,rotating}
\usepackage{amsmath,amssymb}
\usepackage{dsfont}
\restylefloat{table}
\numberwithin{equation}{section}

\newcommand{\vx}{\vec{x}}
\newcommand{\vp}{\vec{p}}

\newcommand{\vk}{\vec{k}}

\newcommand{\be}{\begin{equation}}
\newcommand{\ee}{\end{equation}}
\newcommand{\bea}{\begin{eqnarray}}
\newcommand{\eea}{\end{eqnarray}}

\newcommand{\ket}[1]{|#1\rangle}
\newcommand{\bra}[1]{\langle#1|}

\newcommand{\ta}{\widetilde{A}}
\newcommand{\tb}{\widetilde{B}}
\newcommand{\tf}{\widetilde{f}}
\newcommand{\of}{\overline{f}}

\begin{document}

\title{Is the effective potential, effective for dynamics?.}

\author{Nathan Herring}
\email{nherring@hillsdale.edu} \affiliation{Department of Physics, Hillsdale College, Hillsdale, MI 49242}
\author{Shuyang Cao}
\email{shuyang.cao@pitt.edu} \affiliation{Department of Physics, University of Pittsburgh, Pittsburgh, PA 15260}
 \author{Daniel Boyanovsky}
\email{boyan@pitt.edu} \affiliation{Department of Physics, University of Pittsburgh, Pittsburgh, PA 15260}

\date{\today}

\begin{abstract}
  We critically examine the applicability   of the effective potential within dynamical situations and find, in short, that the answer is negative.  An important caveat of the use of an effective potential in dynamical equations of motion is an explicit violation of energy conservation.
   An \emph{adiabatic} effective potential is introduced
in a consistent quasi-static approximation, and   its   narrow regime of validity is discussed.   Two ubiquitous instances in which even the adiabatic effective potential is not   valid in dynamics are  studied in detail:   parametric amplification in the case of oscillating mean fields, and spinodal   instabilities associated with spontaneous symmetry breaking. In both cases   profuse particle production is directly linked to the failure of the effective potential to describe the dynamics. We introduce a consistent, renormalized,  energy conserving dynamical framework   that is amenable to numerical implementation.   Energy conservation leads to  the emergence of asymptotic highly excited, entangled  stationary states from  the dynamical evolution. As a corollary, decoherence via dephasing of the density matrix in the adiabatic basis is argued to lead to an emergent entropy, formally equivalent to the entanglement entropy.  The results suggest   novel characterization of asymptotic equilibrium states in terms of order parameter vs. energy density.

\end{abstract}

\keywords{}

\maketitle

\section{Introduction}
The effective potential is a very useful concept to study spontaneous symmetry breaking in quantum field theory as originally proposed in refs.\cite{jona,goldstone}. It is defined as the
generating functional of the single particle irreducible Green's functions at \emph{zero four momentum transfer}.  In particular the effective potential informs how radiative corrections modify the symmetry breaking properties of the vacuum\cite{colewein}. While originally the effective potential was obtained by summing an infinite series of Feynman diagrams\cite{colewein}, functional methods\cite{ilio,jackiw,colemanbook,colepoli} provide a systematic and simple derivation in a consistent loop expansion, which has been extended to \emph{equilibrium} finite temperature field theory\cite{dolan,weinberg}. In equilibrium at finite temperature   the effective potential informs on the quantum and thermal  corrections to the free energy landscape as a function of the order parameter, and as such it provides a very useful characterization of phase transitions. The concept of the effective potential plays a fundamental role in cosmology, in particular in the description of possible cosmological phase transitions even during the inflationary era\cite{guth,linde,albrecht,kolb,branrmp}.

An alternative Hamiltonian formulation of the effective potential was advanced in refs.\cite{veff,weinbergwu}, it provides a compelling interpretation of the zero temperature effective potential as the expectation value of the quantum Hamiltonian (divided by the volume) in a coherent state,
in which the (bosonic) field associated with symmetry breaking, namely the order parameter,  acquires a \emph{space-time constant} expectation value (see also \cite{colemanbook,weinbergwu}). The one loop effective potential has also been related to a Gaussian wavefunctional\cite{steven}.

 \textbf{Motivation and  objectives:}

 Although the effective potential was introduced and developed to study \emph{static} aspects of spontaneous symmetry breaking and to identify symmetry breaking minima beyond the classical tree level, it is, however, often implemented in \emph{dynamical} studies   of the time evolution of the   expectation value of the scalar field. Since the effective potential is \emph{defined} for zero four momentum transfer, namely for a static and homogeneous field configuration, the rationale behind its use in a dynamical  situation is the assumption of the validity of some adiabatic approximation. Such assumption ultimately needs scrutiny and justification.

  Our motivation for this study is the ubiquity of the use of the effective potential in dynamical situations in which the expectation value of the scalar field evolves in time. Our objectives are: \textbf{i:)}  to critically examine the validity of using the effective potential in such  dynamical setting, \textbf{ii:)} to  assess the validity of an adiabatic approximation that would justify its use, \textbf{iii:)} identify possible scenarios wherein its use is unjustified, \textbf{iv:)} to  provide an alternative formulation that   overcomes the limitations of its (mis) use, and to study the consequences of the dynamical evolution within this framework.

  In this article we address these aspects at zero temperature in Minkowski space time, obtaining the energy functional and equations of motion including one-loop quantum corrections, which allows us to compare to the one-loop effective potential and exhibit its shortcomings in the simplest case. This study is a prelude towards extending the results both to finite temperature, higher orders and an expanding cosmology in future work.

  \textbf{Brief summary of results:}

  We implement a Hamiltonian approach to obtain the one loop effective potential in the static case and extend it to obtain the energy functional and equations of motion for the expectation value of a scalar field in the dynamical case. An adiabatic effective potential is introduced as a test of whether a quasi static approximation can be reliably applied to the dynamical case, it is explicitly shown that it has a very restricted regime of applicability. Furthermore, we unambiguously show that using the static effective potential in dynamical situations leads to a violation of energy conservation. Two ubiquitous instances are recognized to lead to a breakdown of the adiabatic (quasi static) approximation to the equations of motion: parametric amplification in the case of oscillating mean fields, and spinodal decomposition in the case of spontaneous symmetry breaking. Both phenomena yield profuse particle production which invalidates an adiabatic (quasi-static) approximation and renders the static effective potential an ill-suited description for the dynamics. We introduce a self-consistent, energy conserving, fully renormalized framework to study the dynamical evolution of expectation values of scalar fields. Energy conservation leads us to conjecture the emergence of asymptotic stationary states. These are  characterized by a large occupation number of adiabatic particles in bands, yielding a highly excited entangled state of correlated particle-pairs produced from  resonant transfer of energy from parametric or spinodal instabilities. These highly excited stationary states lead us to suggest a novel characterization of asymptotic equilibrium states in terms of  phase diagrams of \emph{asymptotic order parameter as a function of  energy density}.

  The article is organized as follows: in section (\ref{sec:Veff}) we summarize the Hamiltonian approach to the one loop effective potential in the static case introduced in refs.\cite{veff,weinbergwu}) as a roadmap to extend this formulation to the dynamical case. In section (\ref{sec:adveff}) we extend the Hamiltonian formulation and introduce the framework to study the dynamical case. We also introduce a systematic adiabatic expansion and an adiabatic effective potential and analyze its suitability for describing the dynamics. It is argued that using the static effective potential leads to a violation of energy conservation, and   that the adiabatic effective potential has a very restricted range of validity. In section (\ref{sec:breakdowm}) we study two ubiquitous cases that lead to a breakdown of adiabaticity invalidating the use of the effective potential: \textbf{i:)} parametric amplification when the scalar field oscillates near the minimum of the tree level potential, \textbf{ii:)} spinodal instabilities in the case of spontaneous symmetry breaking. In both cases  we show that parametric and spinodal instabilities lead to profuse particle production which is associated with  the  breakdown of adiabaticity. In section (\ref{sec:hartree}) we introduce a self-consistent, fully renormalized, energy conserving framework to study the dynamical evolution of the expectation value of a scalar field amenable to numerical implementation.  In this section we argue that energy conservation in the dynamics leads us to \emph{conjecture} the emergence of asymptotic stationary, highly excited entangled states from the dynamical evolution with asymptotic values of the order parameter very different from those obtained from an effective potential. In this asymptotic regime, decoherence via dephasing leads to an emergent entropy density
  $$  s = \int \Big[\big(1+ \widetilde{\mathcal{N}}_{\vk}(\infty)\big)\,\ln\big(1+ \widetilde{\mathcal{N}}_{\vk}(\infty)\big)- \widetilde{\mathcal{N}}_{\vk}(\infty))\ln  \widetilde{\mathcal{N}}_{\vk}(\infty)  \Big] \, \frac{d^3k}{(2\pi)^3}\,, $$ where $\widetilde{\mathcal{N}}_{\vk}(\infty)$ is the particle number distribution
as a function of  particle momentum as $t\rightarrow \infty$. This entropy is formally equivalent to an entanglement entropy. Furthermore, we also propose the hitherto unexplored concept of    ``phase diagrams'' of order parameter versus energy density as characterizations of these asymptotic states. Conclusions are summarized in section (VI).

\section{Statics: the  effective potential:  }\label{sec:Veff}

 In this study we focus on one-loop radiative corrections, adopting and extending the formulation of the effective potential of refs.\cite{veff,weinbergwu} which relies on a Hamiltonian description as an alternative to the functional methods, that will be extended to the dynamical case in the next sections.

Let us  consider a real scalar field, $\phi$, in Minkowski spacetime with an action given by
    \be A = \int d^4x \Bigg\{\frac{1}{2} \partial^{\mu}\phi\partial_{\nu}\phi-V(\phi)\Bigg\} \label{Action1}\;,
    \ee
where $V(\phi)$ is the tree level potential. In the interest of generality, we leave this function unspecified at present but consider specific scenarios below from which we draw more general conclusions.

Introducing the   canonical conjugate field momentum operator $ {\pi}(\vx)=\frac{\partial \mathcal{L}}{\partial \phi}=\frac{\partial \phi}{\partial t}$, and upon quantization of the field and its canonical momentum $\phi(x) \rightarrow \hat{\phi}(x), \pi(x) \rightarrow \hat{\pi}(x)$ where the operators $\hat{\phi}(\vx,t); \hat{\pi}(\vx,t)$ obey canonical commutation relations, the field Hamiltonian is given by
    \be H = \int d^3x \Bigg\{\frac{\hat{\pi}^2}{2}+\frac{(\nabla \hat{\phi})^2}{2}+V(\hat{\phi})\Bigg\}\;. \label{Hamiltonian1}\ee

    The Hamiltonian interpretation of the effective potential advanced in refs.\cite{veff,weinbergwu} (see also ref.\cite{colemanbook}) identifies the effective potential as the expectation value of the Hamiltonian operator in a normalized coherent state  $\ket{\Phi}$ in which the field acquires a \emph{space-time independent expectation value}
    \be \varphi = \bra{\Phi}\hat{\phi}(\vx,t)\ket{\Phi}~~;~~\bra{\Phi}\hat{\pi}(\vx,t)\ket{\Phi} =0\,,\label{statexvals}\ee divided by the spatial volume of quantization $\mathcal{V}$,  namely
    \be V_{eff}(\varphi) = \frac{1}{\mathcal{V}}\,\bra{\Phi} {H}\ket{\Phi}\,.\label{veffstatic}\ee We refer to $\varphi$ as a \emph{mean field}, and writing
    \be \hat{\phi}(\vx,t) = \varphi+ \hat{\delta}(\vx,t)~~;~~ \hat{\pi}(\vx,t) \equiv \hat{\pi}_\delta(\vx,t)\,, \label{fieldsplit}\ee the constraints (\ref{statexvals}) imply
    \be \bra{\Phi}\hat{\delta}(\vx,t)\ket{\Phi}=0 ~~;~~\bra{\Phi}\hat{\pi}_\delta(\vx,t)\ket{\Phi} =0\,,\label{statexvalsdel}\ee leading to
    \be V_{eff}  = \,V(\varphi)+ \frac{1}{\mathcal{V}}\, \int d^3x \,\bra{\Phi}\Big\{\frac{\hat{\pi}^2_\delta}{2}+\frac{(\nabla \hat{\delta})^2}{2}+\frac{1}{2}\,\mathcal{M}^2(\varphi)\,\hat{\delta}^2 + \cdots \Big\}\ket{\Phi}\;, \label{Hamiltonian2} \ee where linear terms in $\hat{\delta}$ and $\hat{\pi}_\delta$ vanish by the constraints (\ref{statexvals}), and
    \be \mathcal{M}^2(\varphi) \equiv V''(\varphi)\,. \label{mass} \ee \emph{Assuming} that the effective squared mass $\mathcal{M}^2(\varphi) \geq 0$,  up to quadratic order the Hamiltonian in eqn. (\ref{Hamiltonian2}) describes a free massive field. Hence, we quantize as usual:
    \bea \hat{\delta}(\vx,t)  & = &  \sqrt{\frac{\hbar}{\mathcal{V}}}\,\sum_{\vk} \frac{1}{\sqrt{2\omega_k}}\,\Big[a_{\vk}\, e^{-i\omega_k t}\,e^{i\vk\cdot \vx} + a^\dagger_{\vk}\, e^{i\omega_k t}\,e^{-i\vk\cdot \vx}\Big] \,,\label{quandelta}\\ \hat{\pi}_\delta(\vx,t)  & = &  -i\sqrt{\frac{\hbar}{\mathcal{V}}}\,\sum_{\vk} \frac{\sqrt{\omega_k}}{\sqrt{2}}\,\Big[a_{\vk}\, e^{-i\omega_k t}\,e^{i\vk\cdot \vx} - a^\dagger_{\vk}\, e^{i\omega_k t}\,e^{-i\vk\cdot \vx}\Big] \,,\label{quanpidelta}\eea with
    \be \omega_k(\varphi) = \sqrt{k^2+\mathcal{M}^2(\varphi)} \,. \label{omegak}\ee The constraints (\ref{statexvalsdel}) are implemented by requesting that
    \be a_{\vk}\ket{\Phi} =0 ~,~ \forall \vk \,, \label{anni}\ee in other words, the coherente state $\ket{\Phi}$ is the \emph{vacuum state} for the fluctuations $\hat{\delta}$. In principle, the constraints (\ref{statexvalsdel}) are also fulfilled if  $\ket{\Phi}$ is an eigenstate of the number operator $a^\dagger_{\vk} a_{\vk}$ with eigenvalue $n_k$, however the energy is lowest for the vacuum state with $n_k=0$.

     Taking the infinite volume limit with $\sum_{\vk} \rightarrow \mathcal{V}\,\int d^3k/(2\pi)^3$ and using (\ref{anni}), we find that  the effective potential (\ref{veffstatic}) is given by
    \be V_{eff}(\varphi) = V(\varphi)+ \frac{\hbar}{2} \int \frac{d^3k}{(2\pi)^3}\,\omega_k(\varphi) + \mathcal{O}(\hbar^2) +\cdots \,.\label{Veff1lup}\ee The $\hbar$ in (\ref{Veff1lup}) originates in the $\sqrt{\hbar}$ in the usual field quantization (\ref{quandelta},\ref{quanpidelta}) and implies that the expression (\ref{Veff1lup}) is the \emph{one loop effective potential}. If $\ket{\Phi}$ is an excited eigenstate with $n_k\neq 0$, the integrand in the second term features an extra contribution $n_k\,\omega_k(\varphi)$ thereby rasing the energy.

     That the second term in (\ref{Veff1lup}) is a one-loop contribution is easily understood from the fact that $\bra{\Phi} \hat{\delta}^2(\vx,t)\ket{\Phi}$ is the $\delta$   propagator in the coincidence limit of space-time coordinates, namely the propagator with the end-points joined. The integral is carried out with an ultraviolet cutoff $\Lambda \gg \mathcal{M}(\varphi)$ yielding the one loop effective potential (after setting $\hbar\equiv 1$)
    \be V_{eff}(\varphi) = V(\varphi) + \frac{\Lambda^4}{16\pi^2} + \mathcal{M}^2(\varphi)\,\frac{\Lambda^2}{16\pi^2}-\frac{(\mathcal{M}^2(\varphi))^2}{64\,\pi^2}\,\Big[\ln\Big( \frac{4\Lambda^2}{\mu^2}\Big)-\frac{1}{2}\Big]+\frac{(\mathcal{M}^2(\varphi))^2}{64\,\pi^2}\, \ln\Big( \frac{\mathcal{M}^2(\varphi)}{\mu^2}\Big) \,, \label{veff2}\ee where we have introduced a renormalization scale $\mu$. The ultraviolet divergences must be absorbed into renormalizations of the parameters of the classical potential. Considering the simple example of the tree level potential
    \be V(\varphi) = V_0+\frac{m^2_0}{2}\,\varphi^2+ \frac{\lambda_0}{4}\,\varphi^4 \Rightarrow \mathcal{M}^2(\varphi) = 3\lambda_0\,\varphi^2 + m^2_0 \,, \label{treepot}\ee   introducing the renormalized quantities
    \bea \frac{m^2_R(\mu)}{2} & = & \frac{m^2_0}{2}+ \frac{3\lambda_0}{16\pi^2}\,\Lambda^2-\frac{3\lambda_0}{32\pi^2}\,m^2_0\,\Big[\ln\Big( \frac{4\Lambda^2}{\mu^2}\Big)-\frac{1}{2}\Big]\,\label{mren}\\
\frac{\lambda_R(\mu)}{4} & = & \frac{\lambda_0}{4}- \frac{9\,\lambda^2_0}{32\pi^2}\, \Big[\ln\Big( \frac{4\Lambda^2}{\mu^2}\Big)-\frac{1}{2}\Big] \,,\label{lambdaren}\\
V_{0R}(\mu) & = & V_0 + \frac{\Lambda^4}{16\pi^2}+m^2_0\, \frac{\Lambda^2}{16\pi^2}-\frac{m^4_0}{64\pi^2}\,\Big[\ln\Big( \frac{4\Lambda^2}{\mu^2}\Big)-\frac{1}{2}\Big] \,, \label{voren} \eea and replacing bare by renormalized quantities up to one loop, the renormalized effective potential becomes
\be V_{effR}(\varphi;\mu) = V_{0R}(\mu)+\frac{m^2_R(\mu)}{2}\,\varphi^2+\frac{\lambda_R(\mu)}{4}\,\varphi^4+ \frac{(\mathcal{M}^2_R(\varphi))^2}{64\,\pi^2}\, \ln\Big( \frac{\mathcal{M}^2_R(\varphi)}{\mu^2}\Big) \,. \label{veffren}\ee The effective potential is independent of the renormalization scale $\mu$ which has been introduced to
render the logarithms dimensionless, therefore it obeys the renormalization group equation\cite{colewein}
 \be \mu  \frac{d}{d\mu} V_{effR}(\varphi;\mu) =0 \,.  \ee

\subsection{Fermionic contributions: Yukawa interactions}\label{subsec:yukawa}
The Hamiltonian framework for the effective potential also lends itself straightforwardly to include the contribution from fermions. Consider for example, massless Dirac fermions Yukawa coupled to the scalar field $\phi$ with Lagrangian density

\be \mathcal{L}_f = \overline{\psi}\Big( i{\not\!{\partial}}- Y\,\phi \Big)\psi \,.\label{fermion}\ee Performing the shift $\hat{\phi}(\vx,t) = \varphi + \hat{\delta}(\vx,t)$ the Dirac Hamiltonian becomes to leading order
\be H_f = \int d^3 x \,\psi^\dagger \Big(i \vec{\alpha}\cdot \nabla + m_f(\varphi)\Big)\psi \,,\label{Hf}\ee where the effective Dirac fermion mass is
\be m_f(\varphi) = Y\varphi\,,\label{mf}\ee and
we neglected the interaction term $Y\,\hat{\delta}\, \psi^\dagger   \psi$ as it yields higher order loop corrections to the effective potential.
Quantization now is straightforward in terms of creation and annihilation of particles and antiparticles and the usual Dirac spinor wave functions: positive and negative frequency solutions of the Dirac equation with a mass $m_f(\varphi)$. The state $\ket{\Phi}$ now corresponds to the fermion vacuum and the scalar boson coherent state, yielding the following fermionic contribution to the effective potential
\be V^{(f)}_{eff}(\varphi) = -2 \int \omega^{(f)}_k(\varphi) \,\frac{d^3k}{(2\pi^3)}~~;~~\omega^{(f)}_k(\varphi) = \sqrt{k^2+m^2_f(\varphi)}\,.\label{vfereff}\ee Introducing an upper momentum cutoff $\Lambda$, a calculation similar to the one for the bosonic case yields the fermionic contribution to the effective potential
\be  V^{(f)}_{eff}(\varphi) = -  \Bigg[  \frac{\Lambda^4}{4\pi^2} + m^2_f(\varphi)\,\frac{\Lambda^2}{4\pi^2}-\frac{m^4_f(\varphi) }{16\,\pi^2}\,\ln\Big( \frac{4\Lambda^2}{\mu^2}\Big)+\frac{m^4_f(\varphi)}{16\,\pi^2}\,\ln\Big( \frac{m^2_f(\varphi)}{\mu^2}\Big)\Bigg]\,.  \label{vferef}\ee Renormalization proceeds as in the bosonic case.
These results are in agreement with those of refs.\cite{veff,weinbergwu,colemanbook}, and while these are fairly well known, the main objective of re-deriving them here within the Hamiltonian formulation is to highlight the following aspects:\textbf{ i)} the effective potential is a \emph{static} quantity, \textbf{ii)} it can be directly obtained from the Hamiltonian framework as the expectation value of the quantized Hamiltonian in the  particular coherent state $\ket{\Phi}$ yielding the expectation values (\ref{statexvals}). \textbf{iii)} This analysis informs on the renormalization aspects associated with the effective potential and serve as a guide to the renormalization in the dynamical case studied in the next sections.

We will not pursue the fermionic case further in this article, postponing its detailed study to a forthcoming article. The main and only reason for introducing the case of Yukawa coupling to Fermions is to highlight that the Hamiltonian formulation of the effective potential reproduces the well known results obtained by summation of Feynman diagrams or functional methods which are best suited for the static case and is not restricted to the bosonic case.

Although the effective potential is a static quantity, it is often used in effective equations of motion for $\varphi$, namely
\be \ddot{\varphi}(t)+ \frac{d}{d\varphi} V_{eff}(\varphi(t)) =0 \,,\label{eqcon}\ee or in cosmology including the Hubble-friction term\cite{kolb}. Underlying this use of the \emph{static} effective potential in a dynamical equation of motion is the un-spelled (and unexamined) assumption of quasi-static or adiabatic evolution, namely that the evolution of $\varphi(t)$ is ``slow enough'' that using a static effective potential is warranted.

A main objective of this work is to critically assess this assumption, identify under which circumstances it is warranted, analyze the circumstances when it is not and provide a consistent framework to study the dynamics.

\section{Dynamics: an adiabatic effective potential?}\label{sec:adveff}

When $\varphi$ evolves in time, the dynamics must be studied by evolving a density matrix in time, for which the Schwinger-Keldysh or in-in formulation is better suited\cite{schwinger,keldysh,maha,beilok,boyadiss}. We here provide an alternative by extending to the dynamical case, the Hamiltonian formulation of the effective potential up to one loop advanced in refs.\cite{veff,weinbergwu}  and summarized in the previous section (see also ref.\cite{colemanbook}). In the  dynamical situation   the constraints  (\ref{statexvals}) are relaxed allowing the homogeneous expectation values of field and canonical momentum  to depend on time.

Therefore, we consider a coherent state $\ket{\Phi}$ such that the field operator $\hat{\phi}$  and its canonical conjugate momentum $\hat{\pi}$ acquire   spatially homogeneous but time dependent expectation values, namely
    \be \bra{\Phi}\hat{\phi}(\vec{x},t)\ket{\Phi} = \varphi(t) ~~;~~ \bra{\Phi}\hat{\pi}(\vec{x},t)\ket{\Phi} = \dot{\varphi}(t)\;, \label{classical_condition}\ee
where $\varphi(t)$ is a \emph{classical} homogeneous field, namely a \emph{dynamical mean field}. Therefore $\ket{\Phi}$ characterizes a spatially translational invariant coherent state (annihilated by the spatial momentum operator). To describe this dynamical case, we work in the Heisenberg picture wherein operators evolve in time but states do not, hence the coherent state $\ket{\Phi}$ is time independent.  The Heisenberg field equations obtained from the action (\ref{Action1}) are
      \be  \partial^2_t \hat{\phi}-\nabla^2 \hat{\phi}+V'(\hat{\phi})   =0 \label{EoM1}\;, \ee
with $\frac{\partial}{\partial \phi} \equiv ^{'}$,   which are obviously also satisfied as expectation values in the time independent coherent state $\ket{\Phi}$, namely
\be \bra{\Phi}\Big[ \partial^2_t \hat{\phi}-\nabla^2 \hat{\phi}+V'(\hat{\phi})\Big] \ket{\Phi}  =0 \label{EoMeval}\;, \ee
and we  consider the following initial conditions,
    \bea \bra{\Phi}\hat{\phi}(\vec{x},0)\ket{\Phi} =\varphi(0) \\
        \bra{\Phi}\hat{\pi}(\vec{x},0)\ket{\Phi} = \dot{\varphi}(0) \,. \eea
As in the static case we  write the field operators separating  the ``classical" expectation values, namely the mean fields,   and the quantum fluctuations,
    \be \hat{\phi}(\vec{x},t) = \varphi(t)+ \hat{\delta}(\vec{x},t)~~;~~  \hat{\pi}(\vec{x},t) = \dot{\varphi}(t)+ \hat{\pi}_{\delta}(\vec{x},t) \label{Class_Qtm }\;,\ee
which in accordance with equation (\ref{classical_condition}) requires vanishing expectation values of the fluctuations in the coherent state $\ket{\Phi}$, namely
    \be \bra{\Phi}\hat{\delta}(\vec{x},t)\ket{\Phi} = 0 ~~;~~ \bra{\Phi}\hat{\pi}_{\delta}(\vec{x},t)\ket{\Phi}=0\label{Fluc_EV}\;. \ee

Using equations (\ref{Class_Qtm }) and (\ref{Fluc_EV}) the expectation value of the field Hamiltonian operator (\ref{Hamiltonian1}) can be written as
\be \bra{\Phi}\hat{H}\ket{\Phi} = \mathcal{V}\,\Big[ \frac{\dot{\varphi}^2(t)}{2}+ V(\varphi(t))\Big] + \bra{\Phi}H_\delta\ket{\Phi} \label{HamiltonianEV}\;, \ee
with
\be H_{\delta} = \int d^3x \Bigg\{\frac{\hat{\pi}^2_{\delta}}{2}+\frac{(\nabla \hat{\delta})^2}{2} + \frac{V''(\varphi(t))}{2}\hat{\delta}^2+  \cdots\Bigg\}\,, \label{hdeldef}\ee
where   the expection values of the linear terms in $\hat{\pi}_\delta,\hat{\delta}$ vanish by eqn. (\ref{Fluc_EV}),    $\mathcal{V}$ is the spatial volume in which the field is quantized and we have expanded the potential around the mean field $\varphi(t)$. The Heisenberg equation of motion (\ref{EoM1}) becomes
\be  \ddot{\varphi}(t)+ V'(\varphi(t))+ \partial^2_t \hat{\delta}-\nabla^2 \hat{\delta}+V^{''}(\varphi(t))\,\hat{\delta}+
\frac{1}{2}\, V^{'''}(\varphi(t))\hat{\delta}^{\,2} + \cdots    =0 \,,\label{EoM2}  \ee and similarly with its expectation value in the coherent state $\ket{\Phi}$ (\ref{EoMeval}). A related approach has also been  considered to explore dynamical aspects in ref.\cite{coher}.

\vspace{1cm}

\subsection{Quantization}\label{subsec:quantization}

 The quadratic terms in $\hat{\delta}$ in the Hamiltonian (\ref{hdeldef}) describe a free field theory but now with a \emph{time dependent mass term} $V''(\varphi(t))$. Therefore, in analogy with the static case, we proceed to quantize the theory by considering the solutions of the \emph{linearized} equations of motion, describing a free field with a time dependent mass $V''(\varphi(t))$, namely
    \be \partial^2_t \hat{\delta}-\nabla^2\hat{\delta}+V''(\varphi(t))\hat{\delta}=0\,.\label{linearHeis}  \ee The field operators  $\hat{\delta}(\vx,t)~;~\hat{\pi}_{\delta}$  are expanded in Fourier modes in the   quantization volume  $\mathcal{V}$,
   \bea \hat{\delta}(\vx,t)  & =  & \frac{\sqrt{\hbar}}{\sqrt{\mathcal{V}}}\,\sum_{\vk} \Big[ a_{\vk}\,g_k(t)\,e^{i\vk\cdot\vx} + a^\dagger_{\vk}\,g^*_k(t)\,e^{-i\vk\cdot\vx} \Big]\,,\label{expadelta}\\ \hat{\pi}_\delta(\vx,t) & = & \frac{\sqrt{\hbar}}{\sqrt{\mathcal{V}}}\,\sum_{\vk} \Big[ a_{\vk}\,\dot{g}_k(t)\,e^{i\vk\cdot\vx} + a^\dagger_{\vk}\,\dot{g}^*_k(t)\,e^{-i\vk\cdot\vx} \Big]\,,\label{expapidelta}
   \eea
   and the mode functions, $g_k(t)$, obey the equation of motion
    \be \ddot{g}_k(t) +\omega^2_k(t) g_k(t) = 0\;\; ; \; \omega^2_k(t) \equiv \big[k^2+V''(\varphi(t))\big]  \label{ModeTimeEvo}\;, \ee
with the Wronskian condition dictated by canonical commutation relations to be
    \be \dot{g}_k(t) g^{*}_k(t) -g_k(t)\dot{g}^{*}_k(t)=-i\,. \label{wronsk} \ee The annihilation and creation operators $a_{\vk},a^\dagger_{\vk}$  are time independent because the mode functions $g_k(t)$  are solutions of the mode equations (\ref{ModeTimeEvo}), thereby the fluctuation field $\hat{\delta}(\vx,t)$ is a solution of the linearized Heisenberg field equation (\ref{linearHeis}). They  obey standard canonical commutation relations and the condition
    \be a_{\vk}\ket{\Phi} =0   \,,\label{coherentcond}\ee hence ensuring the fulfillment of the conditions (\ref{Fluc_EV}).
    Just as in the static case, the conditions (\ref{Fluc_EV}) are also fulfilled if the state $\ket{\Phi}$ is an eigenstate of the number operator $a^\dagger_{\vk}a_{\vk}$ with eigenvalue $n_k$.
    We have explicitly included $\sqrt{\hbar}$ in the expressions (\ref{expadelta},\ref{expapidelta}) to highlight below the connection  with the loop expansion\cite{jackiw,colemanbook,weinberg} as in the static case of the previous section. We can now obtain the energy density and the expectation value of the Heisenberg field equation, with
    \be \bra{\Phi}H_\delta\ket{\Phi} = \frac{\hbar}{2}\sum_{\vk} \Big[ |\dot{g}_k(t)|^2 + \omega^2(t)\,|g_k(t)|^2 \Big]+\mathcal{O}(\hbar^2) \,,\label{AdiabaticHamiltonian1}\ee we obtain up to $\mathcal{O}(\hbar)$ (one loop)
    \be \mathcal{E}=
    \frac{\bra{\Phi}\hat{H}\ket{\Phi}}{\mathcal{V}} =   \frac{1}{2}\,{\dot{\varphi}^2(t)}+ V(\varphi(t)) +\mathcal{E}_f(t),, \label{enerdens}\ee  where we have introduced the energy density from one loop quantum fluctuations
    \be \mathcal{E}_f(t) = \frac{\hbar}{2} \int \frac{d^3k}{(2\pi)^3}\,\Big[ |\dot{g}_k(t)|^2 + \omega^2(t)\,|g_k(t)|^2 \Big] \,.\label{enerdensfluc}\ee If the state $\ket{\Phi}$ is an eigenstate of the number operator with eigenvalue $n_k$ the bracket in the above expression is multiplied by $1+2n_k$, just as in the static case this state would be of higher energy. The vacuum state with $n_k=0$ yields the lower fluctuation energy in the static and the dynamical cases.

    Similarly, up to one loop order ($\mathcal{O}(\hbar)$) the expectation value of the Heisenberg field equation (\ref{EoMeval}) in the coherent state $\ket{\Phi}$  becomes
    \be  \ddot{\varphi}(t)+ V'(\varphi(t)) +
\frac{\hbar}{2}\, V^{'''}(\varphi(t))\,\int \frac{d^3k}{(2\pi)^3}\, |g_k(t)|^2       =0 \,,\label{EoM2exval}  \ee to obtain both expressions we used the linearized equations of motion (\ref{linearHeis}), the field expansions (\ref{expadelta},\ref{expapidelta}), the constraint (\ref{coherentcond}) and the infinite volume limit $\sum_{\vk} \rightarrow \mathcal{V} \int d^3k/(2\pi)^3$.

The  $\mathcal{O}(\hbar)$ terms  in (\ref{enerdens},\ref{EoM2exval}) are \emph{one loop} contributions, these arise from $\bra{\Phi} \hat{\pi}^2_\delta \ket{\Phi};\bra{\Phi} \hat{\delta}^{\,2}  \ket{\Phi}$, which are simply the propagators (or derivatives) closed onto themselves. Solving the Heisenberg field equations, along with the constraints (\ref{Fluc_EV})  in a systematic perturbative expansion in the non-linearities, will generate higher orders in the loop expansion. In this article we focus on the one loop ($\mathcal{O}(\hbar)$) contribution to the energy density and equations of motion of the mean field.

The total Hamiltonian does not depend explicitly on time, hence energy is conserved and in the Heisenberg picture the state $\ket{\Phi}$ is time independent, therefore the expectation value of the energy density in the coherent state $\ket{\Phi}$ is conserved, namely $\dot{\mathcal{E}} =0$.  Using the equations of motion of the mode functions (\ref{ModeTimeEvo}) and the form of the time dependent frequencies (\ref{ModeTimeEvo}), it is straightforward to find
\be \dot{\mathcal{E}} = \dot{\varphi}(t) \Big[ \ddot{\varphi}(t)+ V'(\varphi(t)) +
\frac{\hbar}{2}\, V^{'''}(\varphi(t))\,\int \frac{d^3k}{(2\pi)^3}\, |g_k(t)|^2    \Big] = 0 \,, \label{conserene}\ee therefore the expectation value of the equation of motion (\ref{EoM2exval}) is the statement of conservation of the (expectation value) of the energy density.

This dynamical conservation law is of paramount importance: if the amplitude of the modes $g_k(t)$ grows in time the fluctuation contribution to the energy density grows at the
expense of the \emph{classical} part of the  energy, resulting in a \emph{damping} of the $\varphi(t)$ amplitude. As it will be studied in detail below, growth of $|g_k(t)|$ is a consequence of instabilities and particle production. Therefore instabilities in the fluctuations entail \emph{dissipative damping}\cite{boyadiss} of $\varphi(t)$. In turn, as discussed in detail below, these instabilities entail the breakdown of a quasi-static or adiabatic approximation and imply that using the static effective potential in the equation of motion of the mean field  is unwarranted.

An important corollary of this analysis is that replacing the second and  third terms in the equation  of motion (\ref{EoM2exval}) by the field derivative of the static effective potential in the case when $\varphi(t)$ evolves in time clearly \emph{violates energy conservation}. This is because energy is conserved only when the mode functions $g_k(t)$ are the solutions of the mode equations (\ref{ModeTimeEvo}) and not of the form $e^{\mp i\omega_k t}$ as used in the calculation of the static effective potential as is explicit in the quantization (\ref{quandelta}, \ref{quanpidelta}) for the static case. This observation will become more clear with the analysis in the next section.

\subsection{Adiabatic Approximation.}\label{subsec:adappx}
Using the effective potential in the equations of motion of the mean field is usually argued to describe the dynamics in a \emph{quasi-static} or adiabatic approximation. Here we introduce the adiabatic expansion that consistently implements this approximation to understand its regime of validity.
 Given the time-dependence of the frequencies in equation (\ref{ModeTimeEvo}), we seek an approximate solution for the mode functions  in terms of a Wentzel-Kramers-Brillouin (WKB) ansatz\cite{birrell}
    \be g_{k}(t) = \frac{e^{-i\int_0^t\,W_k(t')dt'}}{\sqrt{2W_k(t)}}\,,\label{adiadef} \ee
which when inserted into equation (\ref{ModeTimeEvo}) reveals that $W_k(t)$ must satisfy
    \be W^2_k(t) = \omega^2_k(t)-\frac{1}{2}\Bigg[ \frac{\ddot{W}_k}{W_k} -\frac{3}{2}\frac{\dot{W}_k^2}{W_k^2}\Bigg]\,. \ee
The resulting equation can be solved in an \emph{adiabatic expansion}
    \be W_k^2(t) = \omega_k^2(t) \Bigg[1-\frac{1}{2}\frac{\ddot{\omega}_k}{\omega_k^3}+\frac{3}{4}\Big(\frac{\dot{\omega_k}}{\omega_k^2}\Big)^2+\cdots \Bigg]\,. \label{adiaexp1}\ee
In such an expansion, terms which contain $n$-derivatives of $\omega_k$ are known as of n-th order adiabatic. Inspecting the resulting equation reveals that it contains exclusively terms of even adiabatic order.

Using the WKB ansatz and assuming that $W_k(t)$ is real, one can show that
\begin{align}
    |g_k(t)|^2 &= \frac{1}{2W_k(t)} \label{g_squared} \\
    |\dot{g}_k(t)|^2 &= \frac{W_k(t)}{2}\Bigg [1+\frac{1}{4}\Big(\frac{\dot{W}_k}{W_k^2}\Big)^2\,\Bigg]\;, \label{gdotsquared}
\end{align}
which can be combined with equation (\ref{AdiabaticHamiltonian1})  to give
    \be \bra{\Phi} \hat{H}_{\delta} \ket{\Phi} = \frac{1}{4}\sum_k\Bigg\{W_k(t)\Bigg [1+\frac{1}{4}\Big(\frac{\dot{W}_k}{W_k^2}\Big)^2\,\Bigg] + \frac{\omega^2_k}{W_k(t)} \Bigg\}\,. \label{exHdel3} \ee
We now proceed by invoking the adiabatic expansion, equation (\ref{adiaexp1}), and expanding this expectation value up to \emph{2nd order adiabatic}. After carrying out these algebraic manipulations we obtain up to second adiabatic order
    \be \bra{\Phi} \hat{H}_{\delta} \ket{\Phi}  =  \frac{1}{2}\, \sum_k \omega_k\,\Bigg\{ 1+\frac{1}{8}\Big(\frac{\dot{\omega}_k}{\omega_k^2}\Big)^2 + \cdots\Bigg\}\,, \label{AdiabaticHamiltonian2}\ee
    \be |g_k(t)|^2 = \frac{1}{2\omega_k(t)}\, \Bigg[1+\frac{1}{4}\frac{\ddot{\omega}_k}{\omega_k^3}-\frac{3}{8}\Big(\frac{\dot{\omega_k}}{\omega_k^2}\Big)^2+\cdots \Bigg] \,,\label{gk2ads}\ee

    where the dots stand for terms of   higher adiabatic order.

    Following the analysis of the static case one \emph{may} introduce an \emph{adiabatic effective potential} as
    \be V^{(ad)}_{eff}(\varphi) \equiv V(\varphi) + \frac{1}{\mathcal{V}}\,\bra{\Phi}\hat{H}_{\delta}\ket{\Phi}\,.\label{Veffad}\ee

With the result (\ref{AdiabaticHamiltonian2}), we can now express this adiabatic effective potential up to 2nd adiabatic order, obtaining ($\hbar=1$)
    \be V^{(ad)}_{eff}(\varphi)  \equiv  V(\varphi(t)) +\frac{1}{2} \int \frac{d^3k}{(2\pi)^3}\;\omega_k(t) +\frac{1}{16} \int \frac{d^3k}{(2\pi)^3}\;\frac{\dot{\omega}_k^2(t)}{\omega_k^3(t)}\,. \label{effective_potential2}  \ee
  Recalling the definition of the frequencies, $\omega_k(t)$, given by equation (\ref{ModeTimeEvo}), (\ref{effective_potential2}) becomes
    \be  V^{(ad)}_{eff}(\varphi)  \equiv V(\varphi(t)) +\frac{1}{2}\int \frac{d^3k}{(2\pi)^3} \sqrt{k^2 +V''(\varphi(t))}+\frac{\dot{\varphi}^2(t)}{64}(V'''(\varphi(t)))^2\int\frac{d^3k}{(2\pi)^3}\frac{1}{(k^2+V''(\varphi(t)))^{5/2}}\,. \label{effective_potential3} \ee
The identification of this expression with an  \emph{adiabatic effective potential}   warrants discussion. The first term represents the usual classical potential energy density of the field configuration. The second term is a zeroth order adiabatic correction which encodes the effects of the quantum fluctuations. Notice this term is identical to the usual result for the one-loop effective potential (\ref{Veff1lup}) found in section (\ref{sec:Veff}) for the static case,  but now in terms of the dynamical expectation value $\varphi(t)$. This is of course expected because the zeroth order adiabatic does not include any terms with time derivatives of $\varphi(t)$. This term features all the ultraviolet divergences found within the context of the static effective potential (\ref{veff2}) and  would underpin using the usual effective potential in the evolution equation for $\varphi(t)$ as in eqn. (\ref{eqcon}).

 However, the third term represents the second order adiabatic correction which is a consequence  of quantum fluctuations. This term is a distinct consequence of the time-dependence of the expectation value, $\varphi(t)$, and is completely missed if one assumes that the usual form of the effective potential extends without qualification to the scenario of a \textit{dynamical expectation value} as in eqn.(\ref{eqcon}).

The integral expression for the second adiabatic order correction can be evaluated in a straightforward manner provided we  \emph{assume} $V''(\varphi) > 0$:
    \be \frac{\dot{\varphi}^2}{64}(V'''(\varphi(t)))^2\int\frac{d^3k}{(2\pi)^3}\frac{1}{(k^2+V''(\varphi(t)))^{5/2}} = \frac{\dot{\varphi}^2}{384\pi^2}\frac{(V'''(\varphi(t)))^2}{V''(\varphi(t))}\;\; ;\;  \Big(V''(\varphi(t))>0\Big)\;. \label{2ndOrder_Adia_effective_potential}\ee

It is noteworthy that this contribution (and the higher adiabatic orders) is ultraviolet finite, albeit it may feature infrared divergences whenever $V''(\varphi(t))$ vanishes, signalling the  breakdown of the adiabatic approximation.

Of course,   there are additional, higher adiabatic order corrections to the effective potential which at and beyond second adiabatic order all feature time derivatives of $\varphi(t)$ and they are all ultraviolet finite. At present, we restrict ourselves to a study of the second order adiabatic correction, which suffices to highlight if and when the adiabatic approximation breaks down.

    \vspace{1mm}

\subsection{Equations of motion and the adiabatic effective potential.}\label{subsec:EOM}

 In the scenario where the expectation value of the scalar field is time-dependent, $\bra{\Phi}\hat{\phi}(\vec{x},t)\ket{\Phi} = \varphi(t)$, we are interested in the dynamics of this ``classical" field. Inserting equations (\ref{Class_Qtm }) and (\ref{Fluc_EV}) into the expectation value of the Heisenberg equations of motion for $\hat{\phi}$, equation (\ref{EoM1}), and expanding up to $\mathcal{O}(\delta^2)\propto \hbar$ yields the following equation of motion for the expectation value,
    \be \ddot{\varphi} +V'(\varphi) +\frac{1}{2}V'''(\varphi)\bra{\Phi}\hat{\delta}^2(\vec{x},t)\ket{\Phi} =0\,, \ee
which upon using the Fourier expansion for the fluctuation given by (\ref{expadelta}), and upon setting $\hbar \equiv 1$, becomes
    \be   \ddot{\varphi}+ U'(\varphi) =0 \;,\label{EoMU}   \ee where we have \emph{defined}
    \be  U'(\varphi) \equiv V'(\varphi) +\frac{1}{2}V'''(\varphi)\,\int \frac{d^3k}{(2\pi)^3}\, |g_{k}(t)|^2 \,.\label{Uprime} \ee

  The important question is,  does $U'= \frac{\partial U}{\partial \phi}= \frac{\partial V^{(ad)}_{\text{eff}}}{\partial \phi}\,?$ with $V^{(ad)}_{eff}(\varphi)$ given by equation (\ref{Veffad}), which up to second adiabatic order is given by (\ref{effective_potential2},\ref{effective_potential3}).

To investigate the relationship between   $U'$, and $dV^{(ad)}_{eff}(\varphi)/d\varphi$, we begin by using the result of the WKB ansatz, (\ref{g_squared}), and the adiabatic expansion, (\ref{adiaexp1}), to obtain $U'$ up to second order adiabatic:
\bea &U'(\varphi) = V'(\varphi) +\frac{1}{2}V'''(\varphi)\, \int \frac{d^3k}{(2\pi)^3}\,\frac{1}{2W_k} \\
 &U'(\varphi) \simeq V'(\varphi) +\frac{1}{4}V'''(\varphi)\, \int \frac{d^3k}{(2\pi)^3}\,\Bigg[ \frac{1}{\omega_k} + \frac{1}{4} \,\frac{\ddot{\omega}_k}{\omega_k^4}-\frac{3}{8} \,\frac{\dot{\omega}_k^2}{\omega_k^5} +\dots \Bigg]\,. \eea
For comparison, using equation (\ref{effective_potential2}), we can obtain $dV^{(ad)}_{\text{eff}}/d\varphi$ to second adiabatic order  :
\be \frac{dV^{(ad)}_{eff}}{d\varphi} = V'(\varphi) +\frac{1}{4}V'''(\varphi)  \int \frac{d^3k}{(2\pi)^3} \,\Bigg[ \frac{1}{\omega_k} + \frac{\dot{\varphi}}{4} \,\frac{\dot{\omega}_k}{\omega_k^4}\Big(\frac{V''''}{V'''}-\frac{V'''}{2\omega_k^2}\Big)-\frac{3}{8} \,\frac{\dot{\omega}_k^2}{\omega_k^5} +\dots\Bigg]\,, \label{Veff_prime} \ee
where we have made use of equation (\ref{ModeTimeEvo}) to calculate the necessary derivatives of the frequencies, treating $\varphi$ and $\dot{\varphi}$ independently. Direct comparison of the expressions for $U'$ and $dV^{(ad)}_{{eff}}/d\varphi$ reveals many common terms. However, in the second integral expression lies an apparent discrepancy. Using the definition of the frequencies (\ref{ModeTimeEvo}), we see that
\bea \dot{\omega_k} & = &\frac{\dot{\varphi}}{2\omega_k}V'''\,, \\
\ddot{\omega}_k & = &  \frac{\ddot{\varphi}}{2\omega_k}V'''+\frac{\dot{\varphi}^2}{2\omega_k}V''''-\frac{\dot{\varphi}}{2\omega_k}\frac{\dot{\omega}_k}{\omega_k}V'''\,,\label{hotsad} \eea
and thus
\be \frac{\ddot{\omega}_k}{\omega_k^4}= \frac{\ddot{\varphi}}{2\omega_k^5}V'''+\dot{\varphi}\frac{\dot{\omega}_k}{\omega_k^4}\frac{V''''}{V'''}-\frac{\dot{\varphi}}{2\omega_k^2}\frac{\dot{\omega}_k}{\omega_k^4}V'''\,. \ee
Inserting this result into our expression for $U'(\varphi)$ gives
\be U'(\varphi) = V'(\varphi) +\frac{1}{4}V'''(\varphi)  \int \frac{d^3k}{(2\pi)^3}\,\Bigg[ \frac{1}{\omega_k} +\frac{\dot{\varphi}}{4} \,\frac{\dot{\omega}_k}{\omega_k^4}\Big(\frac{V''''}{V'''}-\frac{V'''}{2\omega_k^2}\Big) +\frac{\ddot{\varphi}}{4} \,\frac{V'''}{2\omega_k^5}-\frac{3}{8} \,\frac{\dot{\omega}_k^2}{\omega_k^5} +\dots \Bigg]\,. \label{U_prime} \ee
Written in this form, we can now manifestly see that $U'$ and $dV^{(ad)}_{{eff}}/d\varphi$ do not match. In particular, using equations (\ref{Veff_prime}) and (\ref{U_prime}),
\be U'(\varphi)-\frac{dV^{(ad)}_{{eff}}(\varphi)}{d\varphi} =  \ddot{\varphi} \,\frac{\big(V'''(\varphi)\big)^2}{16}\,\int \frac{d^3k}{(2\pi)^3}\,\frac{1}{2\omega_k^5}+\cdots =  \ddot{\varphi} \,\frac{\big(V'''(\varphi)\big)^2}{96\,\pi^2\,V''(\varphi)}+\cdots \label{Udiff}\,, \ee where the dots stand for higher derivatives of $\varphi(t)$ and we assumed $V''(\varphi(t))>0$.
Hence, \emph{beyond leading adiabatic order the equation of motion  for $\varphi(t)$ does not involve $dV^{(ad)}_{eff}/d\varphi$ but instead $U'(\varphi)$} defined by eqn. (\ref{Uprime}). Obviously only when time derivatives of the expectation value $\varphi$ vanish, in other words, the \emph{static case},  $U'(\varphi) = dV^{(ad)}/d\varphi$. Therefore, it becomes very clear that while the adiabatic effective potential improves upon the (mis) use of the \emph{static} effective potential in that it includes derivatives of $\varphi(t)$, it is still \emph{not} the proper quantity to use in the equations of motion of $\varphi(t)$.

As stated above   the equation of motion (\ref{EoM2exval}) is tantamount to the statement of the conservation of energy by equation (\ref{conserene}), consequently neglecting the derivatives of $\varphi(t)$ by truncating the adiabatic expansion at some particular order of derivatives  of $\varphi(t)$ entails a violation of energy conservation beyond that order.

A practical question that obviously arises is  the following: if a small violation of energy conservation is tolerated, what would be the range of validity of the adiabatic effective potential in a numerical study of the evolution of $\varphi(t)$ with the equation
\be \ddot{\varphi}(t) + \frac{dV^{(ad)}_{eff}}{d\varphi} = 0\,,  \label{eomvadeff}\ee instead of the exact equation (\ref{EoMU}) with $U'(\varphi)$ defined by (\ref{Uprime})?

For a given classical potential $V(\varphi)$ the result (\ref{Udiff}) yields a quantitative criterion to assess  the regime of validity,   at least up to second adiabatic order. Let us consider first the typical case of
\be V(\varphi) = \frac{1}{2} m^2 \varphi^2 + \frac{\lambda}{4} \varphi^4\,\label{vclasup}  \ee with $m^2>0$ for which
\be U'(\varphi)-\frac{dV^{(ad)}_{{eff}}(\varphi)}{d\varphi} =    \ddot{\varphi}(t)\, \frac{\lambda}{8\pi^2}\,\Bigg[\frac{\Big(3\lambda \varphi^2(t)/m^2\Big)}{1+\Big(3\lambda \varphi^2(t)/m^2\Big)}  \Bigg] \,.\label{dif4}\ee In the small (dimensionless) amplitude regime $3\lambda \varphi^2(t)/m^2 \ll 1$ the difference is \emph{a priori} perturbatively small, the potential (\ref{vclasup}) is dominated by the mass term and the  field oscillates around the minimum $\varphi=0$. This seems to be a regime in which both the adiabatic approximation and the adiabatic potential are reliable, however as we show below in the next section, precisely in this regime there are parametric instabilities resulting in a non-perturbative exponential growth of the mode functions and a complete breakdown of adiabaticity.

In the large amplitude regime $3\lambda \varphi^2(t)/m^2 \gg 1$ the difference (\ref{dif4}) \emph{seems} to be perturbatively small, of $\mathcal{O}(\lambda)$, however, in this regime the adiabatic approximation is no longer reliable for long wavelengths as shown by the following argument. For long wavelengths $k^2 \ll 3\lambda \varphi^2(t)$, and in this large amplitude regime where $V(\varphi) \approx \lambda \varphi^4/4$, the second   order  adiabatic ratio that enters in the adiabatic expansion (\ref{adiaexp1}) becomes
\be \frac{\ddot{\omega}_k(t)}{\omega^3_k(t)} \approx \frac{\ddot{\varphi}(t)}{3\lambda \varphi^3} \,,\label{laramp}\ee  however from the equation of motion at tree level it follows that $\ddot{\varphi}(t) \approx \lambda \varphi^3$ and in this regime we find that
\be \frac{\ddot{\omega}_k(t)}{\omega^3_k(t)} \simeq \mathcal{O}(1) \,,\label{order1}\ee therefore the adiabatic approximation is no longer valid for long wavelength modes with $k^2 \ll   3\lambda \varphi^2(t)$. It is important to highlight that the breakdown of adiabaticity is associated with long wavelength fluctuations, for $k\gg V''(\varphi)$ the adiabatic approximation is reliable, and higher order terms in the adiabatic expansion become further suppressed in this limit.

 This analysis leads us to conclude that the regime of validity of an adiabatic effective potential is severely  restricted to small amplitudes and short times when the parametric instabilities studied in detail in the next section have not yet led to a large growth of the mode functions.

\vspace{1mm}

\section{Breakdown of adiabaticity}\label{sec:breakdowm}
The discussion above highlights that in general the equation of motion cannot be simply written as $\ddot{\varphi}+ V'_{eff}(\varphi)=0$, even in an adiabatic approximation in terms of the adiabatic effective potential, and also illuminates if and when  the adiabatic expansion breaks down. We recognize at least two ubiquitous relevant instances: \textbf{i:)} parametric amplification in the case of oscillating mean fields, \textbf{ii:)} spinodal (tachyonic) instabilities in the case of spontaneous symmetry breaking.

\subsection{Parametric amplification:}

The adiabatic approximation (\ref{adiaexp1}) relies on the assumption that $W^2_k(t) >0$, namely that $W_k(t)$ defined by eqn. (\ref{adiadef}) is real.  This means, for example, that if $V''(\varphi(t))$ is an oscillatory function bounded in time, the resulting mode functions $g_k(t)$ in the adiabatic approximation, given by eqns. (\ref{adiadef},\ref{adiaexp1}) would also be bounded in time, which precludes the possibility of resonances and parametric amplification. Consider the case with tree level potential
\be V(\varphi) = \frac{m^2}{2}\varphi^2 + \frac{\lambda}{4}\varphi^4 \Rightarrow V''(\varphi) = m^2 + 3 \lambda \varphi^2 \,,\label{treeV}\ee with $m^2>0$, and consider that the mean field is oscillating around the minimum of this tree level potential with\footnote{This choice neglects the non-linearities, but will capture the main aspects of parametric amplification. This analysis   also neglects the damping of the amplitude from the back reaction of the fluctuations, which is discussed in detail below. }
\be \varphi(t) = \varphi(0) \cos(mt) \,,\label{oscifi}\ee  defining
\be mt = \tau + \frac{\pi}{2} \,,\label{taudef}\ee the mode equations (\ref{ModeTimeEvo}) become
\be \frac{d^2}{d\tau^2}\,g_k(\tau)+ \big[\eta_k- 2\alpha\,\cos(2\tau)\big]g_k(\tau) =0 \,\label{mathieu} \ee where we introduced the dimensionless variables
\be   \alpha = 3\lambda \frac{\varphi^2(0)}{4\,m^2}~~;~~ \eta = 1+ \kappa^2+2\alpha ~~;~~ \kappa = \frac{k}{m}\,. \label{matvars}\ee The equation (\ref{mathieu}) is recognized as Mathieu's equation\cite{bender,mathieu1,abra,kova}. Floquet's theory\cite{bender} shows that solutions are of the form,
\be g_k(\tau) = e^{i\,\nu_k \tau}\,P_k(\tau)~~;~~ P_k(\tau+\pi) = P_k(\tau)\,\label{floquetsols}\ee where $\nu_k$ is the characteristic exponent of Floquet solutions.  If $\nu_k$ is real the (quasi)-periodic solutions are stable, whereas if $\nu_k$ is complex there is one growing and one (linearly independent) decaying solution. The growing solution is a consequence of the parametric amplification instability associated with resonances, a subject of utmost importance within the theory of cosmological reheating\cite{rehe1,rehe2,rehe3,branreh,rehe4,rehe5,rehe6,yoshi}. The stability of solutions in the $\eta_k-\alpha$ plane have been thoroughly studied in the literature\cite{bender,mathieu1,abra,kova}. Unstable bands emanate from the resonance values $\eta_k = n^2, n=0,1,2 \cdots$ within these bands the characteristic Floquet exponent $\nu_k$ is complex and the mode functions either grow or decay exponentially, the growing mode $g_k(\tau) \propto e^{|\mathrm{Im}\nu_k| \tau}$. For generic initial conditions the general solution is a combination of the growing and decaying solutions. Using the results from refs.\cite{mathieu1,abra,kova} we find that these unstable bands correspond to
\be \kappa^2_{n,-} \leq \kappa^2 \leq \kappa^2_{n,+}~~;~~ \kappa^2 > 0~~;~~ n=0,1,2\cdots  \,. \label{bands} \ee The bands for $n=0,1$ are unphysical because these correspond to negative values of $\kappa^2$, for $n \geq 2$ a power series expansion in $\alpha$ for $\kappa^2_{n,\pm}$ is available, the first few terms (valid for $\alpha \lesssim \mathcal{O}(1)$) are given for
$n=2,3,4$ in appendix (\ref{app:bands}) and displayed in fig. (\ref{fig:bands}).

   \begin{figure}[ht!]
\begin{center}
\includegraphics[height=5in,width=6in,keepaspectratio=true]{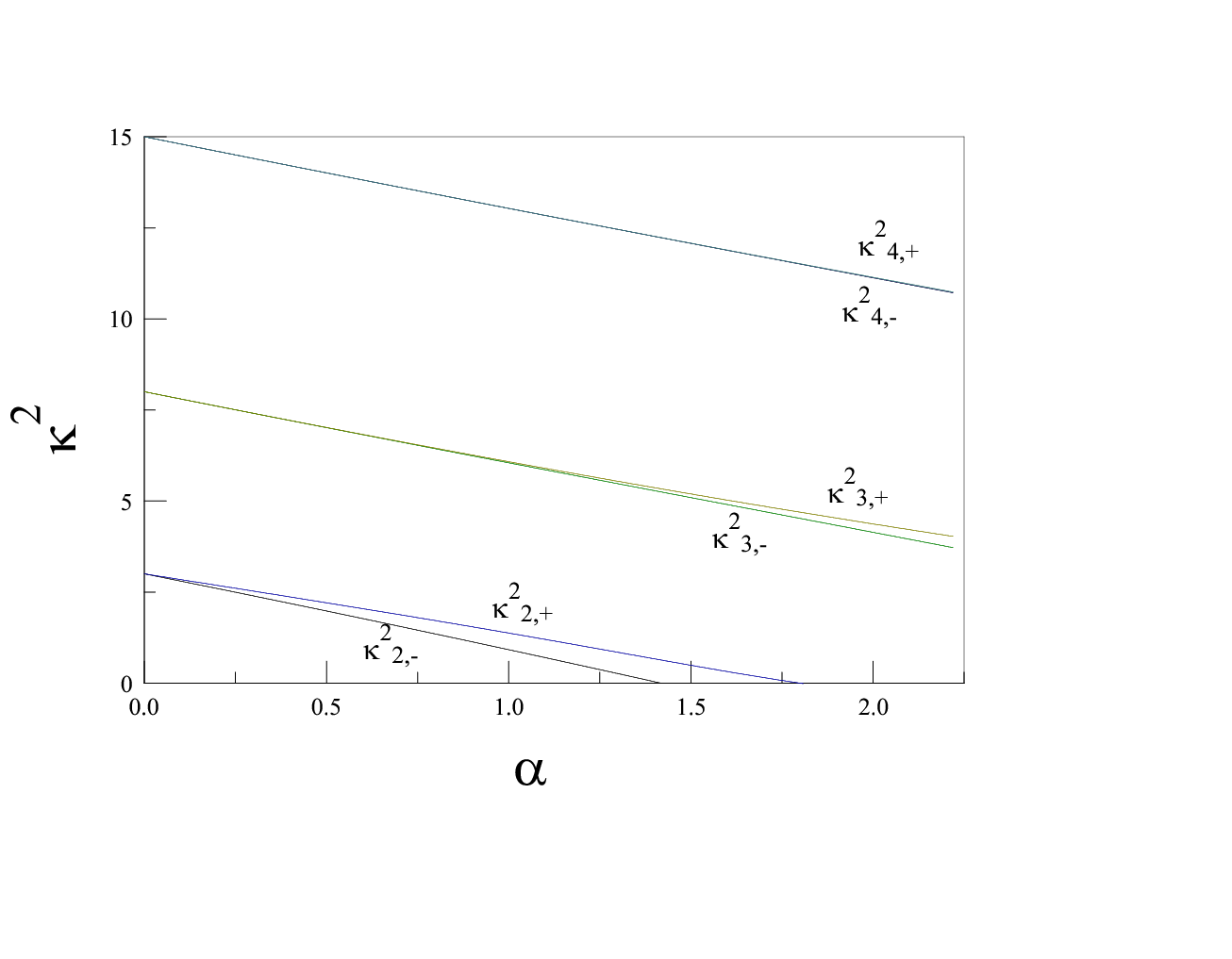}
\caption{Unstable bands for $\kappa^2_{n,-}\leq \kappa^2 = \frac{k^2}{m^2} \leq \kappa^2_{n,+}$ for $n=2,3,4$. The range is constrained by $\kappa^2 >0$. }
\label{fig:bands}
\end{center}
\end{figure}

\begin{figure}[ht!]
\begin{center}
\includegraphics[height=4in,width=3.2in,keepaspectratio=true]{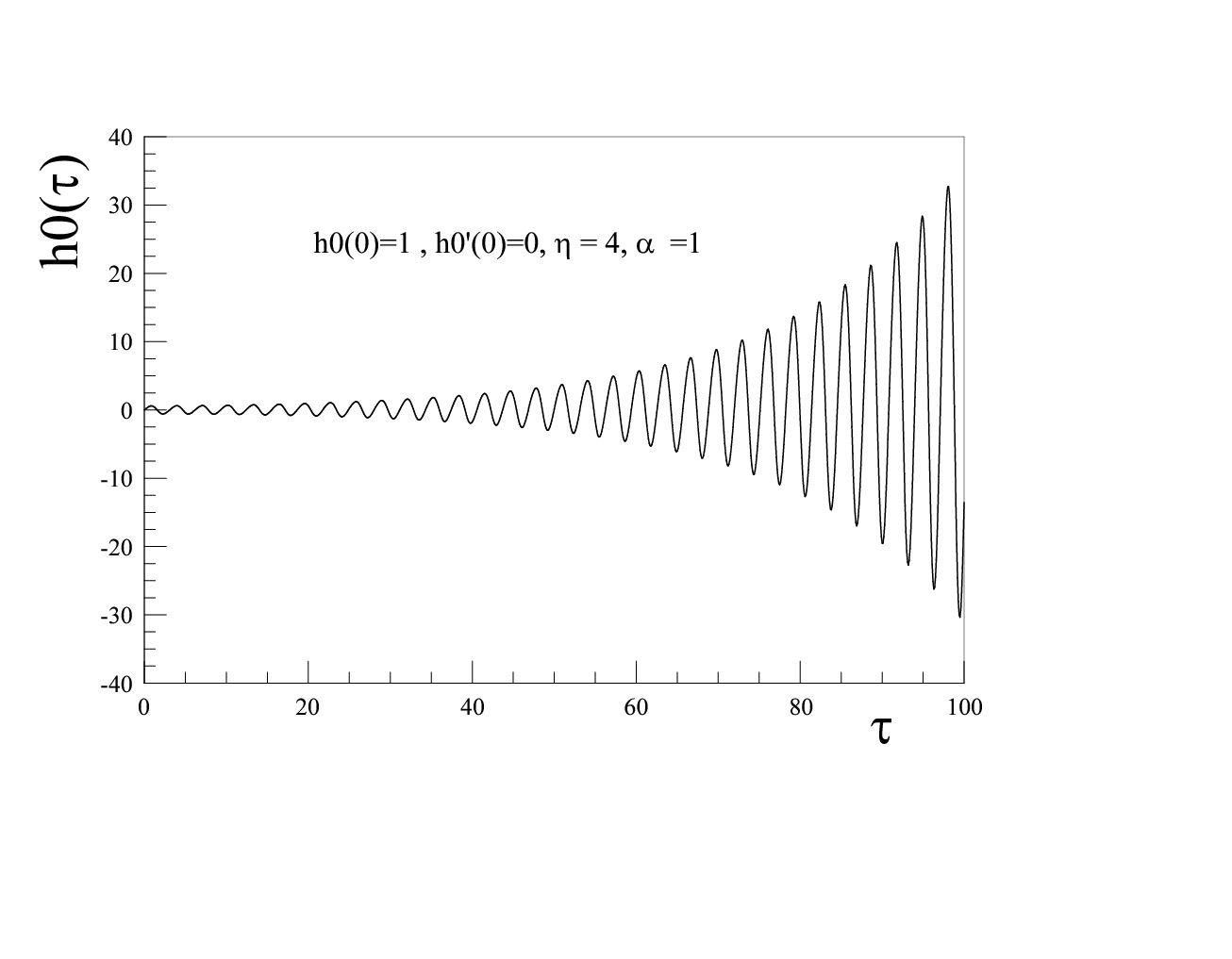}
\includegraphics[height=4in,width=3.2in,keepaspectratio=true]{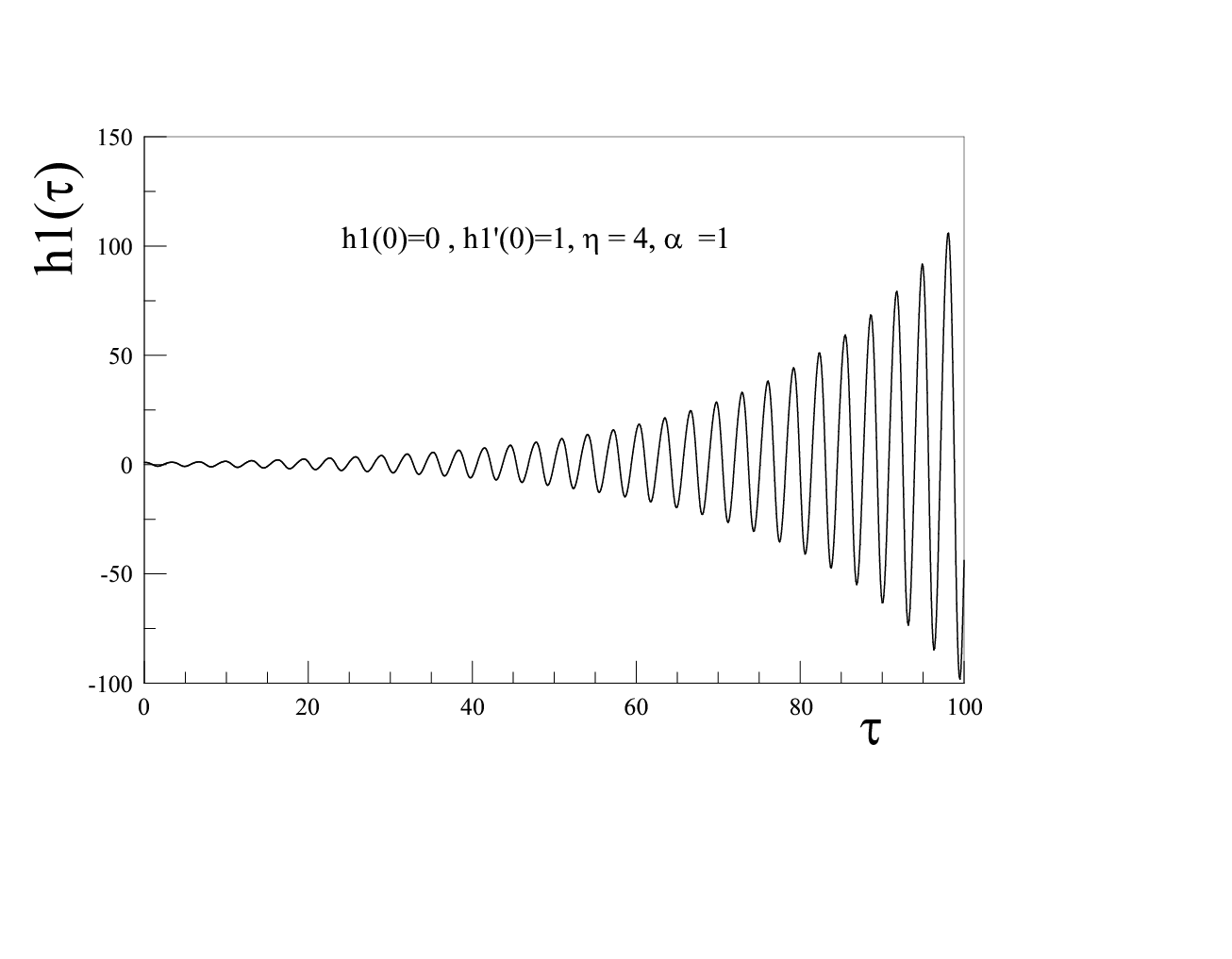}
\caption{Two linearly independent solutions of Mathieu's eqn. (\ref{mathieu}), $h0(\tau),h1(\tau)$ with initial conditions $h0(0)=1,h0'(0)=0;h1(0)=0, h1'(0)=1$, for the   unstable band  for $n=2$, with $\eta = 4$ and $\alpha=1$, corresponding to $\kappa^2=1$, approximately in the middle of the first physical unstable band for $\kappa$. A general solution for a mode function $g_k(\tau)$ is a complex linear combination of $h0(\tau)$ and $h1(\tau)$ satisfying the condition (\ref{wronsk}). }
\label{fig:lisolns}
\end{center}
\end{figure}

Fig. (\ref{fig:lisolns}) shows the numerical evaluation of the  linearly independent solutions $h0(\tau);h1(\tau)$ with initial conditions $h0(0)=0,h0'(0)=1;h1(0)=1,h1'(0)=0$ respectively for the   unstable band with $\eta_k=4;\alpha=1$ corresponding to $\kappa^2=1$, near the middle of the unstable band. This figure clearly shows the exponential growth associated with parametric amplification in the unstable bands. The Floquet exponents may be obtained analytically near the band edges by multi-time scale analysis\cite{bender}, however the actual values of these are not relevant for our general arguments.

\begin{figure}[ht!]
\begin{center}
\includegraphics[height=4in,width=3.2in,keepaspectratio=true]{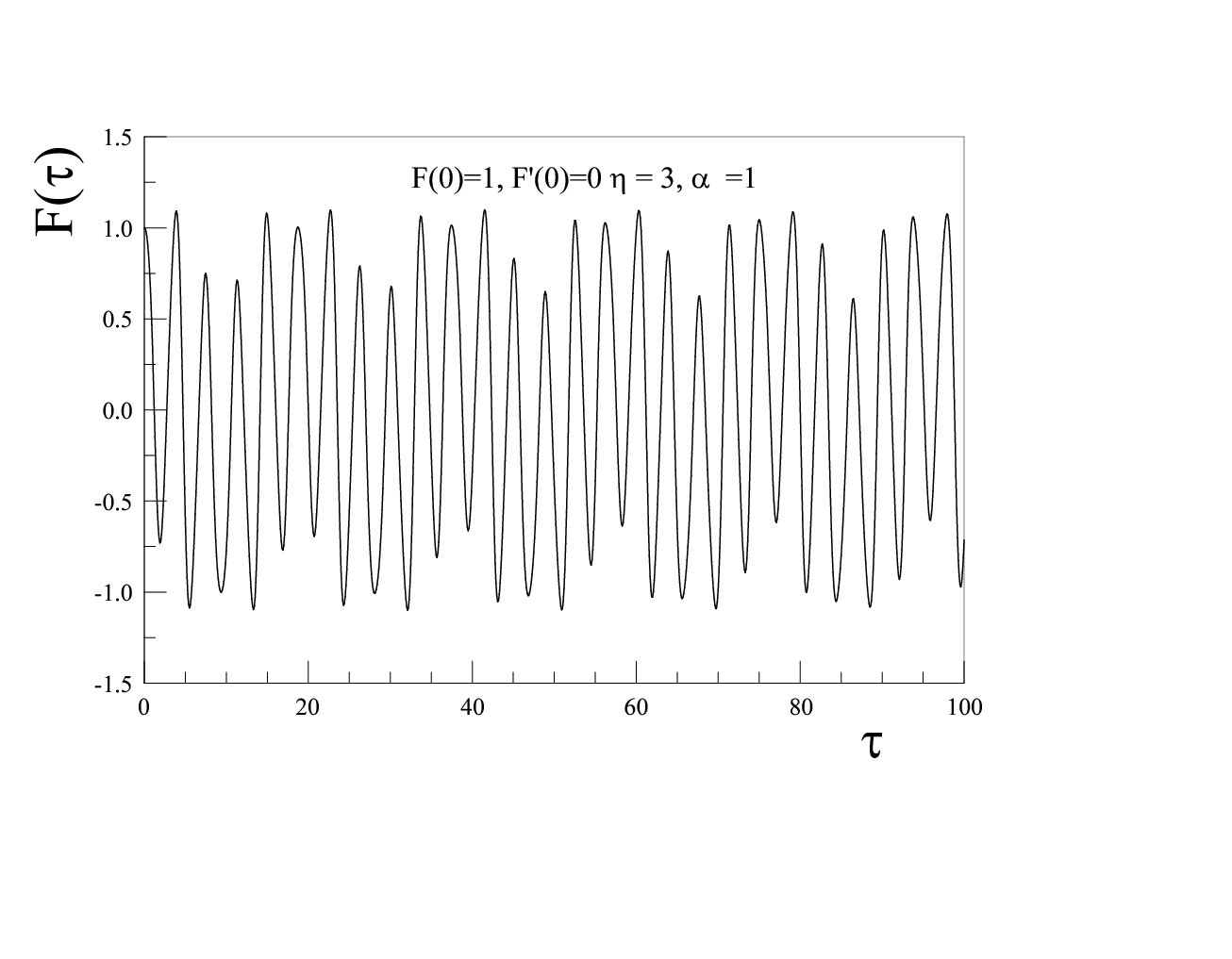}
\includegraphics[height=4in,width=3.2in,keepaspectratio=true]{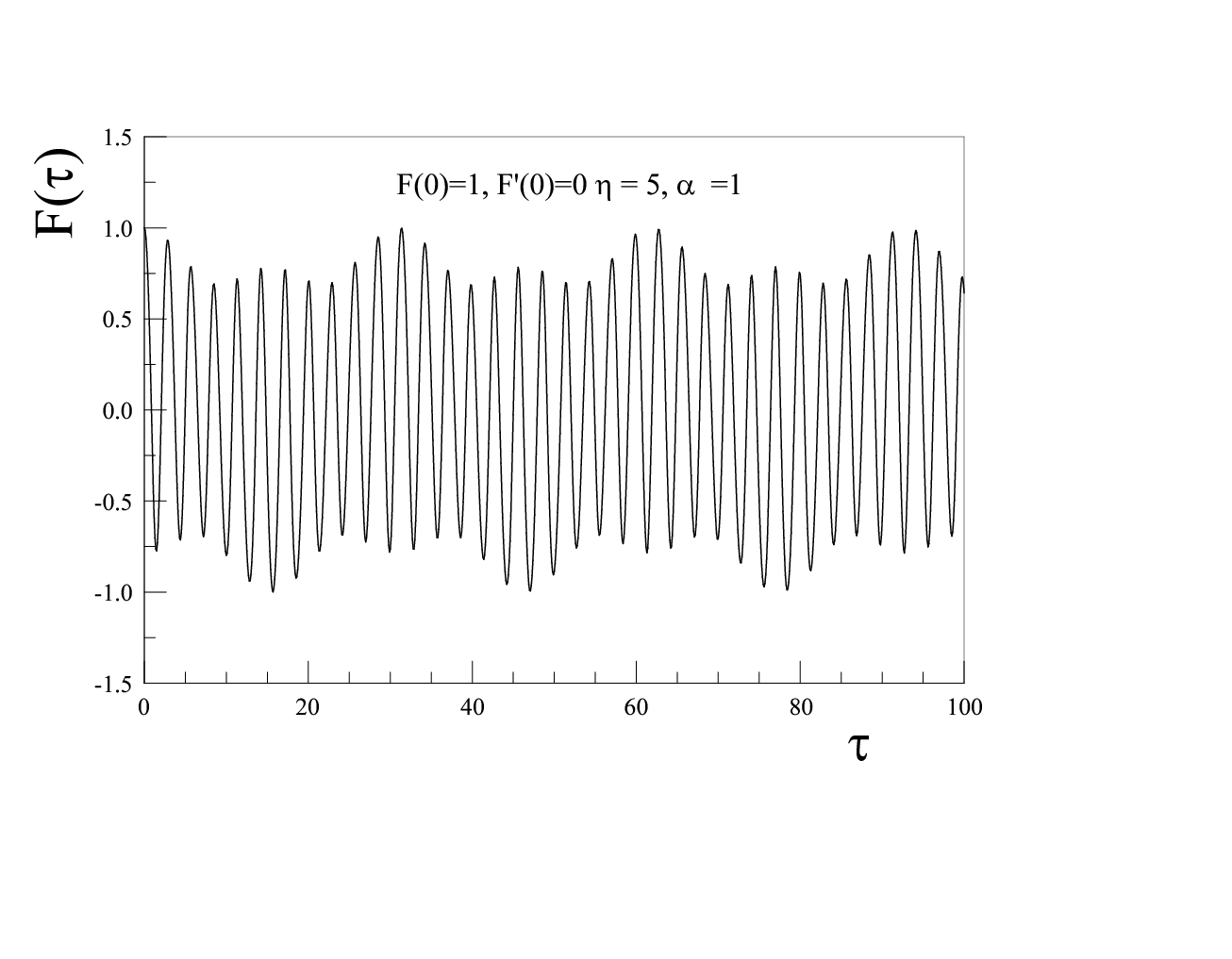}
\caption{Two stable  solutions of Mathieu's eqn. (\ref{mathieu}),  $F(\tau)$ with initial conditions $F(0)=1,F'(0)=0$, for $\eta = 3,5$ and $\alpha=1$ respectively, on either side of the first physical  unstable band at $\eta=4$.  }
\label{fig:stablesolns}
\end{center}
\end{figure}

For comparison, fig. (\ref{fig:stablesolns}) displays the solutions in the stable regions for $\eta=3,5;\alpha=1$,  on either side of the   instability band at $\eta =4$.

The bandwidths $\Delta \kappa^2(n) = \kappa^2_{n,+}- \kappa^2_{n,-} = C_n\alpha^n +\cdots$, with coefficients $C_n$ that become monotonically decreasing with $n$  (see appendix (\ref{app:bands})),  therefore for $\alpha \lesssim \mathcal{O}(1)$ the bands become narrower, as
explicitly shown in fig. (\ref{fig:bands}).

In terms of the momenta $k$, and the amplitude $\varphi(0)$, the bandwidths become
\be \Delta k^2(n) = k^2_{n,+}- k^2_{n,-} = C_n \,\frac{(3\lambda\,\varphi^2(0)/4)^n}{m^{2(n-1)}}  +\cdots \,,\label{bandks}\ee

 This expression highlights that the bands are narrower for weak coupling, large masses or small amplitudes. While this result is  particular to Mathieu's equation, we expect, quite generically that bandwidths for resonances will feature qualitatively similar characteristics as functions of these parameters.

Obviously, the exponential growth with time of the mode functions $g_k(t)$ implies a breakdown of adiabaticity for the values of momentum $k$ within these unstable bands. This can be immediately seen from the adiabatic expansion (\ref{adiaexp1}). Since the frequencies $\omega_k(t)$ are oscillatory, each and all terms in the adiabatic expansion (\ref{adiaexp1}) are oscillatory and bounded in time. Therefore,   $|g_k(t)|^2$ and $|\dot{g}_k(t)|^2$ obtained via the adiabatic approximation (\ref{g_squared},\ref{gdotsquared}) are bounded in time. Instead, the Floquet solutions are unbounded in time for modes within the unstable bands. The unstable Floquet solutions cannot be reliably captured by an adiabatic approximation,   because secular terms associated with resonances\cite{bender} cannot be described  by the adiabatic expansion (\ref{adiaexp1}).

In the fluctuations contribution to the equation of motion (\ref{EoM2exval}) the integral in $k= m\kappa$ sweeps across the unstable bands within which $|g_k(t)|^2$ grows exponentially in time. Consequently the third term in (\ref{EoM2}) grows in time receiving contributions from \emph{all} unstable bands within which there is exponential growth. We emphasize that this
behaviour is not captured by the simple effective potential, nor any adiabatic approximation to it.

The mode equation (\ref{mathieu}) is correct for oscillations of $\varphi(t)$ around an harmonic potential, for anharmonic potentials, the non-linearity induces higher harmonics in the dynamical evolution of $\varphi(t)$, in turn   higher harmonics induce new resonances and unstable bands. However, while the instability chart will be modified by anharmonicity\cite{rehe1,rehe2,boyadiss}, the main observation that the adiabatic approximation cannot reliably describe parametric amplification with the concomitant growth of the mode functions is a generic result of broader significance. This analysis confirms that even in the small amplitude regime when the difference (\ref{dif4}) seems to be perturbatively small, the adiabatic approximation breaks down because of parametric amplification and the adiabatic effective potential is not reliable to describe the dynamics. This analysis of Mathieu's equation, valid for small amplitude,  shows that parametric amplification and exponentially growing modes will continue as long as the amplitude of oscillations is \emph{non-vanishing}. Exponential growth of parametrically amplified modes is effective unless the amplitude of oscillations vanishes.

The breakdown of adiabaticity discussed in section (\ref{subsec:EOM}) and by parametric amplification discussed above is manifest for long wavelengths. For $k^2 \gg \lambda \varphi^2(0)$ the adiabatic ratios $\ddot{\omega}_k(t)/\omega^3_k(t)~;~(\dot{\omega}_k(t)/\omega^2_k(t))^2 \ll 1$ and the width of the unstable bands and the imaginary part of the Floquet exponents become smaller, therefore for large wavevectors the adiabatic approximation is reliable. This is expected on physical grounds as finite amplitude oscillations cannot efficiently transfer energy to very short wavelength modes, in other words, cannot excite high energy degrees of freedom.

\subsection{Spinodal instabilities:}\label{subsec:spino}
The result (\ref{effective_potential3}) for the effective potential up to second adiabatic order, exhibits an important caveat in the case of spontaneous symmetry breaking when the tree level  potential features a maximum implying that   $V''(\varphi)< 0$ in a region $0 \leq |\varphi(t)| \leq |\varphi_s|$ where the actual value of $\varphi_s$ depends on the particular form of the potential. This region   is known as the classical spinodal and corresponds to an unstable region in field space\cite{langer1,langer2,gunton,allen,calzetta,boyaspino,weinbergwu}. In this region the effective mass squared $\mathcal{M}^2(\varphi) \equiv V''(\varphi)$  in eqn. (\ref{mass}) is \emph{negative} and the static effective potential (\ref{veff2}) and its renormalized counterpart (\ref{veffren})  feature an \emph{imaginary part}.  In ref.\cite{weinbergwu} the physical interpretation of  this imaginary part, associated with the spinodal instabilities was elucidated: it yields the lifetime of a quantum state whose wavefunctional is localized in field space within the spinodal region\cite{guthpi}. In refs.\cite{calzetta,boyaspino} the dynamics of such Gaussian wavefunctional and the growth of correlations associated with domain formation  were studied in detail.

  To give a specific example, consider the tree level (classical) potential
\be V(\varphi) = \frac{\lambda}{4}\,\Big(\frac{ \mu^2}{\lambda}-\phi^2\Big)^2 ~~;~~ \mu^2 >0 \,,\label{symbrekpot}\ee within the region
\be  0 \leq \varphi^2 \leq \frac{\mu^2}{3\lambda} \Rightarrow V''(\varphi) <0\,,\label{spinodal} \ee  to which we refer as the (classical) spinodal\cite{langer1,langer2,gunton}, the frequencies $\omega_k$ in eqn. (\ref{ModeTimeEvo}) are given by
\be \omega_k(t) = \sqrt{k^2- |V''(\varphi(t))|}\,.\label{complexwk}\ee  For $k^2 < |V''(\varphi(t))|$ these are purely imaginary describing  the spinodal (tachyonic) instabilities which occur because the field configuration finds itself near  a local maximum of its potential.

 In condensed matter systems these instabilities describe the early stages of a phase transition characterized by the formation of correlated domains, whose typical size, namely the correlation length $\xi(t)$, grows in time\cite{langer1,langer2,gunton}. A similar behaviour emerges in quantum field theory as shown in refs.\cite{weinbergwu,calzetta,boyaspino}, where   the correlation length grows as $\xi(t) \propto \sqrt{t}$ during the early stages, in a similar fashion as in condensed matter systems with a non-conserved order parameter\cite{langer1,langer2,gunton}. These instabilities have also been discussed within the context of inflationary cosmology\cite{guthpi}.

 Since the adiabatic approximation (\ref{adiaexp1}) explicitly requires that $W_k(t)$, introduced in eqn. (\ref{adiadef}), be real-valued, such instabilities characterize a breakdown of adiabaticity.

 This breakdown is explicit in equation (\ref{effective_potential3}) where  both the zeroth and second adiabatic order (the lowest orders) become complex because the momentum integrals receive purely imaginary contributions from the band of unstable wavevectors in the spinodal region $k^2 < |V''(\varphi(t))|$, this is the origin of the imaginary part of the static effective potential in this region. The result (\ref{2ndOrder_Adia_effective_potential})   \emph{assumed} that the frequencies are purely real, namely that $V''(\varphi(t))$ never becomes negative.

 Assuming that $\varphi(t)$ is initially near the maximum of the potential and rolls slowly down the potential hill, at early times  the mode functions  in the band of spinodally unstable momenta are to leading order in an adiabatic (derivative) expansion  neglecting terms with time derivatives of $\varphi(t)$ under the assumption of a ``slow-roll'', are of the form
 \be g_k(t) = r_k\,e^{\int^t_0 \Omega_k(t') dt'} + s_k \,e^{-\int^t_0 \Omega_k(t') dt'}~~;~~ \Omega_k(t) = \sqrt{|V''(\varphi(t))|-k^2}\,,\label{spinomodes}\ee where the complex coefficients $r_k,s_k$ are determined by the initial conditions and Wronskian condition (\ref{wronsk}).    The growth of the mode functions $g_k(t)$ continues until $\varphi(t)$ reaches the inflection or spinodal point $V''(\varphi)=0$ corresponding to the end of the classical spinodal region, beyond which $V''(\varphi(t)) >0$.

The essential conclusion with regards to spinodal instabilities and the effective potential is twofold. \textbf{1:)} If the classical potential features a spinodal region, then a quasi-static, adiabatic description will fail to capture the dynamics of the system above the spinodal point. \textbf{2:)} Moreover, even outside the spinodal region, a significant breakdown of adiabaticity can occur as the spinodal point is approached from below, even when arbitrarily slowly, because the frequencies $\omega_k(t)$ vanish at the spinodal point  and become imaginary above it,  thus rendering a quasi-static, adiabatic approach ineffective.

In a numerical integration of the equations of motion, it is possible to set initial conditions for which $\varphi(t)$ is well below the spinodal and $V''(\varphi) >0$, thereby avoiding the spinodal instabilities altogether. Such setup must also avoid possible excursions of $\varphi(t)$ near the end of the spinodal at which $V''(\varphi(t))=0$ because in this case the adiabatic approximation also breaks down for small momenta. Even restricting initial conditions to avoid the region with $V''(\varphi) \leq 0$, the oscillations of $\varphi(t)$ in the region $V''(\varphi(t))>0$ will lead to parametric instabilities as discussed in the previous section. Therefore insisting  on using the  static effective potential  or even the adiabatic effective potential is  clearly unreliable,   leading to a manifest violation of energy conservation and to completely miss exponentially growing modes associated with  spinodal or  parametric instabilities.

\vspace{1mm}

\subsection{Non-adiabatic particle production: } \label{subsec:energy}

  As emphasized in the above discussion, the equation of motion for $\varphi(t)$, (\ref{EoM2exval}) is the statement of the conservation of the total energy density (\ref{enerdens}) when the mode functions obey the equations (\ref{ModeTimeEvo}). In the case of instabilities, either parametric or spinodal, the  fluctuation contribution to the total energy density, $\mathcal{E}_f(t)$ given by eqn. (\ref{enerdensfluc}), grows at the expense of the first two, ``classical'' terms in the energy density (\ref{enerdens}). In this subsection we seek to establish  a correspondence between the growth of $\mathcal{E}_f(t)$ and particle production.

\vspace{1mm}

\textbf{Parametric instabilities:}

 In the case of parametric instabilities for a convex function $V(\varphi)$ which can always be defined to be positive, the first two terms in (\ref{enerdens}) are manifestly positive and so is the fluctuation term $\mathcal{E}_f(t)$, because $\omega^2_k(t) > 0$. Therefore, energy conservation implies that the non-adiabatic growth of the fluctuation term must result in a damping of the amplitude of $\varphi(t)$. The draining of the ``classical'' part of the energy, namely the first two terms in (\ref{enerdens}) can be interpreted as the profuse production of \emph{adiabatic particles}. This can be understood from the following argument.

 In the expansion of the field in terms of the exact mode functions (\ref{expapidelta}) the annihilation and creation operators $a_{\vk},a^\dagger_{\vk}$ are time independent because the mode functions $g_k(t)$ obey the Heisenberg field equation (\ref{linearHeis}). Following \cite{parker,birrell,ford,fullbook,parkerbook,mukhabook,mottola,dunne} we can introduce \emph{time} dependent operators by expanding in the basis of the zeroth order  adiabatic    particle states. Introducing the zeroth-order adiabatic modes
\be \tf_k(t)= \frac{e^{-i\,\int^{t}\,\omega_k(t')\,dt'}}{\sqrt{2\,\omega_k(t)}} \,,  \label{zerof}\ee we can expand the \emph{exact} mode functions $g_k(t)$ as
\be g_k(t) = \ta_k(t)\,\tf_k(t)+ \tb_k(t)\,\tf^*_k(t) \label{gexpa}\ee and \emph{define} \cite{parker,mottola,dunne}
\be \dot{g}_k(t) = -i \omega_k(t)\,\Big[  \ta_k(t)\,\tf_k(t)-\tb_k(t)\,\tf^*_k(t)\Big] \,. \label{dergexpa}\ee
The relations (\ref{gexpa},\ref{dergexpa}) can be inverted to yield the Bogoliubov coefficients\cite{mottola}
\bea \ta_k(t) & = &  i\,\tf^*_k(t) \Big[\dot{g}_k(t) -i \omega_k(t)  \,g_k(t)  \Big] \label{tilA} \\
\tb_k(t) & = &  -i\,\tf_k(t) \Big[\dot{g}_k(t) +i \omega_k(t) \,g_k(t)  \Big]\,. \label{tilB}\eea

It follows from  the Wronskian condition (\ref{wronsk}) that
\be |\ta_k(t)|^2-|\tb_k(t)|^2 =1\,. \label{Wroab}\ee

 The definition (\ref{gexpa})  yields
\bea a_{\vk}\,g_k(t) + a^\dagger_{-\vk} \, g^*_k (t)  & = & c_{\vk}(t) \,\tf_k(t)+  c^{\dagger}_{-\vk}(t)\,\tf^*_k(t)\,, \label{adiaexp}\\
a_{\vk}\,\dot{g}_k(t) + a^\dagger_{-\vk} \, \dot{g}^*_k (t)  & = & -i\omega_k(t) \,\Big( c_{\vk}(t) \,\tf_k(t)-  c^{\dagger}_{-\vk}(t)\,\tf^*_k(t)\Big)\,, \label{dotadiaexp}\eea where
\be c_{\vk}(t) = a_{\vk}\,\ta_k(t)+a^\dagger_{-\vk}\,\tb^*_k(t)~~;~~c^\dagger_{\vk}(t) = a^\dagger_{\vk}\,\ta^*_k(t)+a_{-\vk}\,\tb_k(t)\,, \label{cops} \ee the condition (\ref{Wroab}) ensures that $c_{\vk}(t);c^\dagger_{\vk}(t)$ obey equal time canonical commutation relations.

Although in principle other definitions of particles are possible, there are two important and compelling aspects that distinguish the zeroth adiabatic basis choice over other possible choices: \textbf{i:)} if there is an asymptotic stationary state such that the frequencies $\omega_k(t) \rightarrow \omega_k(\infty)$ the creation and annihilation operators  become constant in time $c^\dagger(t);c(t) \rightarrow c^\dagger(\infty); c(\infty)$ and the right hand side of (\ref{adiaexp}) describe asymptotic ``out'' states with the time evolution $e^{\mp i \omega_k(\infty) t}$. \textbf{ii:)} the time dependent operators $c_{\vk}(t);c^\dagger_{\vk}(t)$ associated with the zeroth-order adiabatic modes have special significance: it is straightforward to show that the quadratic Hamiltonian $H_{\delta}$ given by eqn. (\ref{hdeldef}) can be written as

\be H_{\delta} =  \sum_{\vk}\hbar\,  \omega_k(t)\Big[ c^\dagger_{\vk}(t) c_{\vk}(t) + \frac{1}{2}\Big]\,,\label{adhdel}\ee

Therefore defining the instantaneous adiabatic vacuum state $\ket{0_a(t)}$ so that
\be c_k(t)\ket{0_a(t)} =0 \, \forall k,t \label{adivac}\ee the Fock states
\be  \ket{n_{\vk}(t)}  = \frac{\Big(c^{\dagger}_{\vk}(t)\Big)^{n_{\vk}}}{\sqrt{n_{\vk}!}}\,  \ket{0_a(t)} ~~;~~ n_{\vk} = 0,1,2 \cdots \,,\label{insta}\ee are instantaneous eigenstates of $H_\delta(t)$   to which we refer as \emph{adiabatic particles}. The number of  adiabatic  particles at a given time  in the coherent state $\ket{\Phi}$ is given by
\be \widetilde{\mathcal{N}}_k (t) = \bra{\Phi}c^\dagger_{\vk}(t)\, c_{\vk}(t)\ket{\Phi} = |\tb_k(t)|^2 \,. \label{adianum}\ee This result can also be understood from the relation (\ref{tilB}) and the Wronskian condition (\ref{wronsk}) which yield
  \be \widetilde{\mathcal{N}}_k (t) = \frac{1}{2\omega_k(t)}\Big[|\dot{g}_k(t)|^2+ \omega^2_k(t) |g_k(t)|^2\Big]- \frac{1}{2}\,, \label{Ngrel}\ee from which it follows that
\be \frac{1}{\mathcal{V}}\bra{\Phi}H_{\delta}(t)\ket{\Phi} = \frac{\hbar}{2} \int \frac{d^3k}{(2\pi)^3}\,\omega_k(t)\Big[1+2\widetilde{\mathcal{N}}_k (t)\Big] \,.\label{exvalHdel}\ee

  Note that if $g_k(t)$ coincides \emph{exactly} with the zeroth order adiabatic order mode function,   then $\ta_k(t)=1;\tb_k(t)=0$ and there is no particle production, however if $g_k(t)$ is a  linear combination of both adiabatic modes $\tf_k(t);\tf^*_k(t)$, the Bogoliubov coefficients $A_k,B_k \neq 0$. This is important, because the zeroth adiabatic order for $g_k(t)$ yields the usual effective potential as shown explicitly above.

  Therefore, we conclude that the failure of    the effective potential to correctly describe the dynamical evolution of $\varphi(t)$ is explicitly a consequence of the    \emph{production of adiabatic particles}. The growth of $g_k(t)$ as a consequence of parametric instabilities leads to profuse particle production. From the relation (\ref{tilB}) it is clear that the exponential growth of $g_k(t)$ within the instability bands yields an exponential growth in the adiabatic particle number.

  The relation of the fluctuation component of the energy density $\mathcal{E}_f(t)$ and particle production can be made explicit from the result (\ref{exvalHdel}), yielding the energy density (\ref{enerdens}) directly in terms of the adiabatic particle number, namely (setting $\hbar=1$)

  \be\mathcal{E}=   \frac{1}{2}\,{\dot{\varphi}^2(t)}+ V(\varphi(t)) +\frac{1}{2} \int \frac{d^3k}{(2\pi)^3}\, \omega_k(t)\, \Big[1 + 2 \widetilde{\mathcal{N}}_k (t) \Big]\,.\label{endenpara} \ee

  Comparing with the one loop static effective potential (\ref{Veff1lup}) we see that the first term in the integral in (\ref{endenpara}) is precisely the one loop
  contribution to the effective potential, now with the mean field $\varphi(t)$ depending on time, therefore we write (\ref{endenpara}) in a more illuminating manner as
   \be\mathcal{E}=   \frac{1}{2}\,{\dot{\varphi}^2(t)}+ V_{eff}(\varphi(t)) +  \int \frac{d^3k}{(2\pi)^3}\, \omega_k(t) \widetilde{\mathcal{N}}_k (t)  \,,\label{endenparavef} \ee
   with
   \be V_{eff}(\varphi(t)) = V(\varphi(t))+\frac{1}{2} \int \frac{d^3k}{(2\pi)^3}\,  \omega_k(t) \,,\label{vefot}\ee being the effective potential extrapolated from the static case (\ref{Veff1lup}) to the dynamical case, given by eqn. (\ref{veff2}), and its renormalized version (\ref{veffren}) with $\varphi \rightarrow \varphi(t)$. The final expression for the energy density (\ref{endenparavef}) shows explicitly that in presence of particle production, the effective potential does not yield the correct description of the dynamics.

   The initial condition on the mode functions
   \be g_k(0)= \frac{1}{\sqrt{2\omega_k(0)}}~~;~~\dot{g}_k(0) = \frac{-i\omega_k(0)}{\sqrt{2\omega_k(0)}}\,,\label{inicondspara}\ee yields
   \be \widetilde{\mathcal{N}}_k (0) =0 \,,\label{Ninipara}\ee corresponding to the zeroth order adiabatic vacuum state. Parametric amplification leads to profuse particle production via the exponential growth of mode functions within the unstable bands with the  concomitant growth of the occupation number of adiabatic particles $\widetilde{\mathcal{N}}_k (t) $.

  Particle production from parametric amplification is a well known phenomenon studied in detail within the context of post-inflationary reheating\cite{rehe1,rehe2,rehe3,branreh,rehe4,rehe5,rehe6,yoshi}. However,  to the best of our knowledge its connection with the shortcomings  of the use of the effective potential to studying the dynamical evolution of the expectation value of a scalar field with radiative corrections has not been previously highlighted.

  \vspace{1mm}

  \textbf{Spinodal instabilities:} If  $|\varphi(t)|< |\varphi_s|$
   spinodal instabilities lead to growth of the mode functions $g_k(t)$ given by eqn. (\ref{spinomodes}) in the band of spinodally unstable modes with $k^2 < |V''(\varphi(t))|$. Because the $\omega^2_k(t)$ are negative   for these  modes,  it is not obvious that the fluctuation contribution to the energy density, namely $\mathcal{E}_f(t)$ given by    eqn.    (\ref{enerdensfluc}),     is positive and grows in time. However the following argument indeed shows that $\dot{\mathcal{E}}_f(t)$ is   positive and grows exponentially:    taking the time derivative of  $\mathcal{E}_f(t)$ and using the mode equations (\ref{ModeTimeEvo}), yields  (setting $\hbar=1$)
   \be \dot{\mathcal{E}}_f(t) = \frac{1}{2}\,\Big(\frac{d}{dt}V''(\varphi(t))\Big)\, \int \frac{d^3k}{(2\pi)^3}\, |g_k(t)|^2 \,,\label{dotef}\ee as $\varphi(t)$ rolls down the potential hill within the spinodal region, $V''(\varphi(t))$ increases as a function of time from a negative value up to $V''(\varphi_s)=0$. Therefore $\dot{\mathcal{E}}_f >0$ and grows exponentially during this regime as a consequence of the  exponential  growth of the mode functions.

   Since the total energy is conserved, the growth in the fluctuation contributions is at the expense of diminishing the ``classical'' part, namely the first two terms in (\ref{enerdens}).

  Obviously there is no possible definition of adiabatic modes within this region as the frequencies are purely imaginary for $k^2< |V''(\varphi(t))|$. Therefore, unlike the case of parametric instabilities discussed above (see eqn. (\ref{endenpara})),  $\mathcal{E}_f(t)$ cannot be written solely in terms of an occupation number of adiabatic particles.   However, as $\varphi(t)$ rolls down the ``hill'' towards a  stable minimum of the potential including radiative corrections, the drain of the ``classical'' part of the energy implies that   its amplitude damps out. The mean field   eventually will oscillate around this minimum below the spinodal point where  the frequencies become  real  $\omega_k(t)=\sqrt{k^2+V''(\varphi(t))}$ with $V''(\varphi(t))>0$. This suggests  us to separate the spinodally unstable modes, for which the maximum unstable wavevector is given by
  \be K_s = |V''(0)| \,,\label{ks} \ee and for $k\leq K_s$ we   define the interpolating  frequencies
  \be \varpi_k(t) = \sqrt{k^2+|V''(\varphi(t))|} \,,\label{omegbs}\ee in terms of which we now introduce the mode functions
  \be \of_k(t) = \frac{e^{-i\int^t\varpi_k(t')dt'}}{\sqrt{2\varpi_k(t)}}\,.  \label{fbs}\ee   Following the steps leading to equations (\ref{gexpa},\ref{dergexpa}), for $k\leq K_s$ we now write

  \bea g_k(t)   & = &    \overline{A}_k(t)\,\of_k(t)+ \overline{B}_k(t)\,\of^*_k(t)\,, \label{gexpabar} \\
  \dot{g}_k(t) &  = &   -i \varpi_k(t)\,\Big[  \overline{A}_k(t)\,\of_k(t)-\overline{B}_k(t)\,\of^*_k(t)\Big] ~~;~~ k \leq K_s \,, \label{dergexpabar}\eea  whereas for $k> K_s$ we use the zeroth-order adiabatic mode functions $\tf_k(t)$ given by (\ref{zerof}) along with the definitions (\ref{gexpa},\ref{dergexpa}).

  The advantage of introducing the  (interpolating) mode functions $\of_k(t)$ and the definitions (\ref{gexpabar},\ref{dergexpabar}) is that we expect that asymptotically at long time, when $\varphi(t)$ oscillates below the spinodal, they merge with  the asymptotic adiabatic modes.

 In analogy with the previous case, for the spinodally unstable wave vectors $k< K_s$  we introduce

\be |\overline{B}_k(t)|^2 \equiv \overline{\mathcal{N}}_k (t) = \frac{1}{2\varpi_k(t)}\Big[|\dot{g}_k(t)|^2+ \varpi^2_k(t) |g_k(t)|^2\Big]- \frac{1}{2}\,. \label{spinoN} \ee

In order to understand particle production within the spinodal region more quantitatively, let us consider an initial condition with $\varphi(t)$ near the (shallow) maximum of the potential and slowly evolving towards the bottom, and set the following initial conditions on the mode functions
\be g_k(0) = \frac{1}{\sqrt{2\varpi_k(0)}}~~;~~ \dot{g}_k(0) = \frac{-i\varpi_k(0)}{\sqrt{2\varpi_k(0)}}\,,\label{inigspino}\ee which fulfill the Wronskian condition (\ref{wronsk}) and yield $\overline{\mathcal{N}}_k (0)=0$, describing the vacuum corresponding to the theory with an ``upright'' harmonic potential with frequencies $\varpi(0)$.

 We can now write $\mathcal{E}_f(t)$ as
  \bea && \mathcal{E}_f(t)   =   \int^{\Lambda}_{0} k^2 \big[\varpi_k(t)\,\Theta(K_s-k)+\omega_k(t)\,\Theta(k-K_s) \big]\,\frac{dk}{4\pi^2} \nonumber \\
 & + & \int^{\Lambda}_{0} k^2 \big[\varpi_k(t)\,\overline{\mathcal{N}}_k(t)\,\Theta(K_s-k)+\omega_k(t)\,\widetilde{\mathcal{N}}_k(t)\,\Theta(k-K_s) \big]\,\frac{dk}{2\pi^2} \nonumber \\ & + &  \Big[V''(\varphi(t))-|V''(\varphi(t))|\Big] \int^{K_s}_0 k^2\, |g_k(t)|^2\,  \frac{dk}{4\pi^2} \,, \label{enefspinod}\eea where $\Lambda$ is an ultraviolet cutoff.

 The total energy density (\ref{enerdens}) becomes
\bea   \mathcal{E} & = &  \frac{1}{2}\,{\dot{\varphi}^2(t)}+   {V}(\varphi(t)) +\int^{\Lambda}_{0} k^2 \big[\varpi_k(t)\,\Theta(K_s-k)+\omega_k(t)\,\Theta(k-K_s) \big]\,\frac{dk}{4\pi^2}  \nonumber \\ & + &  \int^{\Lambda}_{0} k^2 \big[\varpi_k(t)\,\overline{\mathcal{N}}_k(t)\,\Theta(K_s-k)+\omega_k(t)\,\widetilde{\mathcal{N}}_k(t)\,\Theta(k-K_s) \big]\,\frac{dk}{2\pi^2} \nonumber \\
&  + & \Big[V''(\varphi(t))-|V''(\varphi(t))|\Big] \int^{K_s}_0 k^2\, |g_k(t)|^2\,  \frac{dk}{4\pi^2}\,.\label{endenunst} \eea

Although it is not necessary to re-write the energy density in this form because the set of equations (\ref{EoM2exval},\ref{ModeTimeEvo}) contain all the information, there are three important aspects that emerge from eqn. (\ref{endenunst}): \textbf{i:)} although the definition of ``adiabatic particles'' in terms of the mode functions (\ref{fbs}) yielding the number of ``particles'' (\ref{spinoN}) is somewhat arbitrary, any alternative definition will exhibit the growth of such particle number as a consequence of spinodal instabilities. \textbf{ii:)} an advantage of the this definition is that after the mean field begins its oscillations around the broken symmetry minimum below the spinodal point, it follows that $V''(\varphi(t))>0$, therefore $\varpi(t) \rightarrow \omega_k(t)$, and $\overline{\mathcal{N}}_k(t) \rightarrow \widetilde{\mathcal{N}}_k(t)$, namely the definition of the particle number (\ref{spinoN}) thus coincides with the ``adiabatic particle number'', and the last terms in eqns. (\ref{enefspinod},\ref{endenunst}) vanish.  When $\varphi(t)$ begins oscillations around the broken symmetry minimum, namely beyond the spinodal point,   the  evolution of the $g_k(t)$   results in the production of particles by parametric amplification, determined by equation (\ref{adianum}) but now defined in terms of the oscillations around the stable broken symmetry minimum of the tree level potential. Therefore the definition of ``adiabatic modes'' (\ref{fbs}) and particle number (\ref{spinoN}) merge smoothly with the definition of ``adiabatic particles'' within the context of parametric amplification.  Different definitions of `` particle'' are possible, an advantage of the definition in terms of the asymptotic adiabatic mode functions (\ref{fbs}) is that it merges with  the adiabatic modes corresponding to oscillations around stable minima.

This ambiguity notwithstanding, it is clear that spinodal and parametric instabilities, both lead to exponential growth of the exact mode functions $g_k(t)$  which, in turn, leads to profuse particle production. As discussed above, oscillations around a  broken symmetry minimum also lead to parametric amplification and exponential growth of the mode functions, different from the spinodal instability. Therefore in this scenario, particles are profusely produced first during the spinodal state, and when the field is oscillating around the broken symmetry minimum via parametric instability. While the quantitative expression of the number of particles produced depends on the precise definition
of the mode functions $\tf_k(t)$, it is clear that either the zeroth order adiabatic (\ref{zerof}) for parametric   or (\ref{fbs}) for spinodal instabilities, yield profuse particle production as a consequence of either instability. \textbf{iii:)} the last term in the first line in (\ref{endenunst}) features the same ultraviolet divergences as those found to renormalize the effective potential (\ref{veff2}-\ref{voren}). The   last term in (\ref{endenunst}) is  finite, and  it will be argued in the next section that all the terms with occupation numbers are indeed finite, this is certainly the case for the contribution from $\overline{\mathcal{N}}_k(t)$ since only momenta $k \leq K_s$ contribute to these.

\vspace{1mm}

\section{A renormalized,  energy conserving framework:} \label{sec:hartree}

The analysis presented in the previous sections unambiguously point out that the effective potential is not reliable to study the dynamics of the mean field $\varphi(t)$ in a broad range of theories with and without symmetry breaking as a consequence of the various instabilities associated with particle production. Instead, up to one loop (setting $\hbar=1$) the dynamics must be studied by implementing the set of equations
\be  \ddot{\varphi}(t)+ V'(\varphi(t)) +
\frac{1}{2}\, V^{'''}(\varphi(t))\,\int \frac{d^3k}{(2\pi)^3}\, |g_k(t)|^2 =0 \,,\label{finEOMlup}\ee where the mode functions are the solutions of the equations
 \be \ddot{g}_k(t) +\omega^2_k(t) g_k(t) = 0\;\; ; \; \omega^2_k(t) \equiv \big[k^2+V''(\varphi(t))\big]  \label{modes1lup}\;, \ee and fulfill the Wronskian condition (\ref{wronsk}). Complemented with initial conditions on $\varphi(t),\dot{\varphi}(t),g_k(t),\dot{g}_k(t)$ this is closed set of equations with a conserved energy density
\be \mathcal{E}=   \frac{1}{2}\,{\dot{\varphi}^2(t)}+ V(\varphi(t)) +\frac{1}{2} \int \frac{d^3k}{(2\pi)^3}\,\Big[ |\dot{g}_k(t)|^2 + \omega^2(t)\,|g_k(t)|^2 \Big] \,. \label{enerdens1lup}\ee However, as discussed within the context of the static effective potential both (\ref{finEOMlup}) and (\ref{enerdens1lup}) feature ultraviolet divergences that must be absorbed by renormalization of the bare parameters of the theory. The instabilities associated with spinodal decomposition or parametric amplification affect the mode functions for a finite range of momenta $k$: spinodal instabilities only affect mode functions with $k\leq |V''(0)|$, with $|V''(0)|$ the maximum value of $|V''(\varphi)|$ in the spinodal region. Although parametric instabilities affect all values of $k^2$ for which there are resonances that lead to parametric amplification, the bandwidth of the unstable regions becomes smaller for larger values of $k$. On physical grounds, for $k^2 \gg V''(\varphi(0))$ resonant transfer of energy from the ``zero mode'' to high energy modes is inefficient. Furthermore, as analyzed in detail in section (\ref{sec:breakdowm}), the adiabatic approximation fails for low energy, long wavelength  modes: those with $k< K_s \simeq V''(0)$ for spinodal instabilities and those within resonant bands for parametric amplification. However, for $k^2 \gg V''(\varphi(0))$ the adiabatic approximation is valid, in this limit the mode functions
\be g_k(t) \propto \frac{e^{ \pm i k t}}{\sqrt{2k}} \,.\label{hikay}\ee

  The explicit form of the adiabatic effective potential (\ref{effective_potential3}) explicitly shows that the zeroth order adiabatic contribution contains all the ultraviolet divergences and the higher order adiabatic terms are all ultraviolet finite. Furthermore, the analysis leading up to equations (\ref{endenpara},\ref{endenunst}) also clearly shows that   the ``zero point'' contribution $\int d^3k \, \omega_k(t)$ in these expressions  contains the ultraviolet divergences, whereas the occupation number $\widetilde{\mathcal{N}}_k(t)$ are finite since neither spinodal nor parametric instabilities   can excite very high energy modes. As discussed above, in section (\ref{subsec:adappx}) the ``zero point'' contribution is completely determined  by the zeroth adiabatic order of the mode functions $g_k(t)$. Therefore, we separate this ultraviolet divergent contribution by adding it into an effective potential and subtracting it from the fluctuation part by  writing

 \be \mathcal{E}=   \frac{1}{2}\,{\dot{\varphi}^2(t)}+\overline{ V}_{eff}(\varphi(t)) +\mathcal{E}_{fR}(t),, \label{enerdensren} \ee with
 \be   \overline{ V}_{eff}(\varphi(t)) = V(\varphi(t))+  \int^{\Lambda}_{0} k^2  \omega_k(t)\,\Theta(k-k_m) \,\frac{dk}{4\pi^2} \,, \label{veffKm}\ee    and
 \be \mathcal{E}_{fR}(t) = \int^{\Lambda}_{0} \frac{dk}{4\pi^2}k^2 \Big[  |\dot{g}_k(t)|^2 + \omega^2(t)\,|g_k(t)|^2  -  \omega_k(t)\,\Theta(k-k_m) \Big]\,,\label{efren}\ee  is the ultraviolet finite,  renormalized fluctuation contribution to the energy density, where the lower momentum cutoff $k_m$ is given by
 \be k_m =  \bigg\{
               \begin{array}{c}
                 0 \textrm{~~without ~symmetry ~ breaking} \\
                  \sqrt{|V''(0)|} = K_s ~~\textrm{with ~symmetry ~ breaking}\\
               \end{array}
             \,,\label{kmini}\ee to account for the spinodal region in the case of symmetry breaking where the frequencies $\omega_k(t)$ become purely imaginary.

               The integrals of $\omega_k(t)$ are straightforward, for $\Lambda \gg |V''(\varphi(t))|$ we find

 \bea \overline{ V}_{eff}(\varphi) & = &  V(\varphi) + \frac{\Lambda^4}{16\pi^2} + \mathcal{M}^2_R(\varphi)\,\frac{\Lambda^2}{16\pi^2}-\frac{(\mathcal{M}^2_R(\varphi))^2}{64\,\pi^2}\,\Big[\ln\Big( \frac{4\Lambda^2}{\mu^2}\Big)-\frac{1}{2}\Big]+\frac{(\mathcal{M}^2_R(\varphi))^2}{64\,\pi^2}\,\ \ln\Big( \frac{|\mathcal{M}^2_R(\varphi)|}{\mu^2}\Big) \,\nonumber \\
 & - & \Big(\mathcal{M}^2_R(\varphi) \Big)^2\,\mathcal{F}\Bigg[\frac{k_m}{|\mathcal{M}^2_R(\varphi)|^{1/2}} \Bigg]   \,,\label{vefina} \eea
 with
 \bea \mathcal{F}\big[x\big] & = & \frac{1}{32\pi^2} \Bigg\{2\,x\, \Big[x^2+ \mathrm{sign}\big(\mathcal{M}^2_R(\varphi)\big)\Big]^{3/2}  - x\,\mathrm{sign}\big(\mathcal{M}^2_R(\varphi)\big)\,\Big[x^2+ \mathrm{sign}\big(\mathcal{M}^2_R(\varphi)\big)\Big]^{1/2}\nonumber \\ & - &\ln\Big[x+\Big[x^2+ \mathrm{sign}\big(\mathcal{M}^2_R(\varphi)\big)\Big]^{1/2} \Big] \Bigg\}\,,\label{funx}\eea where we have written $V_{eff}(\varphi(t))$ in terms of
 \be  \mathcal{M}^2_R(\varphi) = V^{''}_R(\varphi(t))\,, \ee to compare to the static result (\ref{veff2}).

 Absorbing the ultraviolet divergences in a renormalization of the
 bare parameters of the tree level effective potential at the renormalization scale $\mu$,  and for the case without symmetry breaking, corresponding to $\mathcal{M}^2(\varphi) >0$ with $k_m=0$,  we identify
 \be \overline{V}_{eff}(\varphi(t)) \equiv V^R_{eff}(\varphi(t);\mu)  \,.\label{veffren2}\ee where
 \be V^R_{eff}(\varphi(t);\mu) = V_R(\varphi;\mu)+  \frac{(\mathcal{M}^2_R(\varphi))^2}{64\,\pi^2}\, \ln\Big( \frac{\mathcal{M}^2_R(\varphi)}{\mu^2}\Big)\,\label{Veffren7} \ee is  the renormalized one loop effective potential, with $V_R(\varphi;\mu)$ the renormalized \emph{tree level} potential in terms of the  renormalized parameters.

 In the case when the tree level potential admits symmetry breaking minima and a spinodal region with $\mathcal{M}^2_R(\varphi)<0$, corresponding to the lower momentum cutoff  $k_m=K_s$,  the contribution from the function $\mathcal{F}$ in (\ref{vefina}) excises  the spinodal region with $k^2 < |V''(0)|=K_s$, which of course contributes to the fluctuation part as is explicit in eqn. (\ref{efren}). Since $K_s >\mathcal{M}^2(\varphi)$ it follows that the effective potential $\overline{ V}_{eff}(\varphi)$ defined by eqn. (\ref{vefina}) is real and does not feature the pathologies of the usual effective potential in the spinodal region. It is straightforward to confirm that taking $k_m \rightarrow 0$ for $\mathcal{M}^2(\varphi)<0$ in $\mathcal{F}$ brings back the imaginary part, arising from the logarithm when $\mathrm{sign}(\mathcal{M}^2(\varphi))< 0$.

  For the case of tree level potential (\ref{treepot}), the renormalization proceeds exactly as in equations (\ref{mren}-\ref{voren}) yielding equation (\ref{veffren}) for the first line of (\ref{vefina}).

 The equation of motion for the mean field (\ref{finEOMlup}), can be similarly written as a fully renormalized equation. To achieve this, again we add and subtract the contribution from the zero adiabatic order, re-writing (\ref{finEOMlup}) as
 \be \ddot{\varphi}(t)+ V^{'}_R(\varphi(t)) +  V^{'''}_R(\varphi(t))\,\int^{\Lambda}_{0}  k^2\, \frac{\Theta(k-k_m)}{2\omega_k(t)}\,\frac{dk}{4\pi^2} \, + \,
  V^{'''}_R(\varphi(t))\,\int^{\Lambda}_0 \frac{dk}{4\pi^2}\, k^2\Bigg[|g_k(t)|^2-\frac{\Theta(k-k_m)}{2\omega_k(t)}\Bigg] =0 \,,\label{eomfiren}  \ee from which we recognize that
  \be  V^{'}_R(\varphi(t)) +   V^{'''}_R(\varphi(t))\,\int^{\Lambda}_{0}  k^2\, \frac{\Theta(k-k_m)}{2\omega_k(t)}\,\frac{dk}{4\pi^2}  = \frac{d}{d\varphi} \overline{V}^R_{eff}(\varphi;\mu) \,,\label{deriV}\ee with $\overline{V}^R_{eff}(\varphi;\mu)$ given by eqns. (\ref{veffKm},\ref{vefina}) after absorbing the ultraviolet divergences into renormalization of the bare parameters at the renormalization scale $\mu$. We can now write the energy density and equation of motion for the mean field and mode functions (up to one loop) in a manifestly energy conserving (since we added and subtracted the ultraviolet divergent contributions) and   fully renormalized form
  \bea \mathcal{E} & = &   \frac{1}{2}\,{\dot{\varphi}^2(t)}+ \overline{V}^R_{eff}(\varphi(t);\mu) +\mathcal{E}_{fR}(t)\,, \label{enerdensren2} \\
   & &  \ddot{\varphi}(t) + \frac{d}{d\varphi} \overline{V}^R_{eff}(\varphi;\mu)  \, + \,
  V^{'''}_R(\varphi(t))\,\int^{\Lambda}_0 \frac{dk}{4\pi^2}\, k^2\Bigg[|g_k(t)|^2-\frac{\Theta(k-k_m)}{2\omega_k(t)}\Bigg] =0 \,,\label{eomfiren2}\\
   & &\ddot{g}_k(t) +\omega^2_k(t) g_k(t) = 0\;\; ; \; \omega^2_k(t) \equiv \big[k^2+V''_R(\varphi(t))\big]\,,  \label{ModeTimeEvoren}
  \eea
   with $\overline{V}^R_{eff}(\varphi;\mu)$ is the renormalized effective potential defined by eqn. (\ref{veffKm}) where the ultraviolet divergences have been absorbed into a renormalization of the bare parameters of the tree level potential at the renormalization scale $\mu$, and $V_R(\varphi(t))$ is the \emph{tree level potential} in terms of renormalized parameters. The renormalized fluctuation contributions $\mathcal{E}_{fR}(t)$, given by eqn. (\ref{efren}) and the last term in (\ref{eomfiren2}) are ultraviolet finite and account for all of the particle production processes resulting from spinodal and parametric instabilities.

   \textbf{Initialization:} The set of equations (\ref{eomfiren2},\ref{ModeTimeEvoren}) form a self-consistent,   energy conserving closed set of equations that describe an initial value problem amenable to numerical implementation, upon appending initial conditions on the mean field and mode functions. The initial conditions on the mean field are simple:
   \be \varphi(t=0)\equiv \varphi(0)~;~\dot{\varphi}(t=0) \equiv \dot{\varphi}(0) \,,\label{mfini}\ee
   those of the mode functions are subject to the Wronskian condition (\ref{wronsk}) and depend on whether the mean field initially is within the spinodal region or outside it.

   \textbf{i:) $V''_R(\varphi(0)) > 0:$} in this case all modes can be initialized as
   \be g_k(0) = \frac{1}{\sqrt{2\,\omega_k(0)}}~~;~~ \dot{g}_k(0) = \frac{-i\,\omega_k(0)}{\sqrt{2\,\omega_k(0)}}~~;~~ \omega_k(0) = \sqrt{k^2+V''_R(\varphi(0))}\,.\label{inigs1} \ee This initial condition implies that
   the adiabatic number $\widetilde{\mathcal{N}}_k(0) = 0$, and is compatible with the renormalization procedure described above because
   \be |\dot{g}_k(0)|^2+\omega^2_k(0)\,|g_k(0)|^2 = \omega_k(0) \,,\label{iniene}\ee therefore the renormalized energy density from fluctuations in eqn. (\ref{efren}) is ultraviolet finite  initially and the renormalization of ultraviolet divergences is the same as during the time evolution,  regardless of whether the (renormalized) tree level potential features symmetry breaking or not.

    \textbf{ii:) $V''_R(\varphi(0)) < 0:$} in this case the renormalized tree level potential features symmetry breaking minima and a spinodal region. If $\varphi(0)$ is within the spinodal region, a suitable set of initial conditions are
    \bea g_k(0)  & = &  \Bigg\{ \begin{array}{c}
                                 \frac{1}{\sqrt{2\varpi_k(0)}}~~\mathrm{for}~~k^2\leq|V''_R(\varphi(0))| \\
                                  \frac{1}{\sqrt{2\omega_k(0)}}~~\mathrm{for}~~ k^2>|V''_R(\varphi(0))|
                               \end{array} \,,\label{gkaspi}\\
       \dot{g}_k(0)  & = &  \Bigg\{ \begin{array}{c}
                                 \frac{-i\varpi_k(0)}{\sqrt{2\varpi_k(0)}}~~ \mathrm{for}~~ k^2\leq |V''_R(\varphi(0))| \\
                                  \frac{-i\omega_k(0)}{\sqrt{2\omega_k(0)}}~~ \mathrm{for}~~ k^2>|V''_R(\varphi(0))|
                               \end{array} \,,\label{dotgkaspi}    \eea    with $\varpi_k(t) = \sqrt{k^2+|V''_R(\varphi(0))|}$.    These initial conditions imply that the interpolating and adiabatic particle numbers $\overline{\mathcal{N}}_k(0)=0;\widetilde{\mathcal{N}}_k(0)=0$. Furthermore,    at $t=0$ the integrand in eqn. (\ref{efren}) vanishes  identically for $k>k_m$,    yielding an ultraviolet finite renormalized energy density of fluctuations at  all times, including at $t=0$. Therefore, this set of initial conditions is explicitly compatible with the renormalization procedure, because the ultraviolet divergences at the initial time are renormalized in the same manner as the ultraviolet divergences at any other time during the time evolution.

                               Although different initial conditions for the mode functions, subject to the Wronskian conditions (\ref{wronsk}) may be chosen, the compatibility with the renormalization procedure described in the previous section must be carefully assessed for alternative initial conditions. The set above is fully compatible with the renormalization procedure, thereby guaranteeing that there are no new ultraviolet divergences associated with the initial value problem\cite{baacke} and that the renormalization framework is consistent all throughout the time evolution, namely the same counterterms remove the ultraviolet divergences at the initial and at any later time.

                               The  set of renormalized equations (\ref{eomfiren2}-\ref{ModeTimeEvoren}) along with the initial conditions (\ref{mfini}-\ref{dotgkaspi}) thus describes completely a self-consistent  initial value problem which is manifestly energy conserving and fully consistent with
                               the renormalization prescription at all times that is amenable to straightforward numerical implementation.

\subsection{Consequences of energy conservation: asymptotic stationary fixed points?:}\label{subsec:enercons}

 Energy conservation entails that instabilities must eventually shut-off since exponential growth of fluctuations cannot continue indefinitely. Particle production via instabilities combined with energy conservation leads us to the \emph{conjecture} of emerging asymptotic highly excited stationary states as fixed points of the dynamical evolution described by the closed set of equations (\ref{enerdensren2}-\ref{ModeTimeEvoren}).  Both spinodal and parametric instabilities must shut-off asymptotically as a consequence of energy conservation, implying that $\varphi(t)$ is below the spinodal and must approach a constant because any oscillatory behavior results in parametric instabilities, however small
 the amplitude of the oscillation. Therefore asymptotically  $\varphi(t) \rightarrow \varphi(\infty)$ with $\varphi(\infty)$ a constant so that $V^{''}(\varphi(\infty)) >0$. Therefore,  it follows that $\omega_k(t) \rightarrow \omega_k(\infty)$ and the mode functions $g_k(t)$ approach the asymptotic solution
 \be g_k(t) \rightarrow \frac{1}{\sqrt{2\omega_k(\infty)}}\,\Big[\alpha_k\,e^{-i\omega_k(\infty)t} + \beta_k \, e^{i\omega_k(\infty)t}\Big] \,.\label{asygks}\ee The relations (\ref{tilA},\ref{tilB}) yield in this asymptotic limit
 \be   \ta_k(t) \rightarrow      \alpha_k \,e^{i\gamma_A} ~~;~~  \tb_k(t)   \rightarrow      \beta_k \,e^{i\gamma_B}\,,\label{ABasym}
 \ee   with $\gamma_{A,B}$ constant phases, and from  (\ref{cops})  it also follows that
 \be c_k(t) \rightarrow c_k(\infty)~~;~~ c^\dagger_k(t) \rightarrow c^\dagger_k(\infty)\,,\label{copsasy}\ee hence the annihilation and creation operators of the instantaneous zero adiabatic order Fock states become constant.
 To understand clearly the underpinnings of this conjecture   let us consider separately the cases without and with  spontaneous symmetry breaking.

 \vspace{1mm}

 \textbf{i:) Without symmetry breaking:} let us focus on the case of the   simple tree level potential (\ref{treeV}) (with renormalized parameters) as a paradigmatic example,  and an initial condition on $\varphi(0),\dot{\varphi}(0)$ allowing for large amplitude oscillations around the minimum of the tree level potential at $\varphi =0$. With $\mathcal{M}^2(\varphi) > 0$ and $k_m=0$ the contribution from the function $\mathcal{F}$ in (\ref{vefina}) vanishes and $\overline{V}^R_{eff} = V^R_{eff}$, the one loop effective potential (see eqn. (\ref{veffren2})).

The total energy density is conserved and the mode functions obey the equations (\ref{ModeTimeEvoren}), although for large amplitudes the analysis based on Mathieu's equation is no longer valid, we   still expect resonances leading to instability bands within which the mode functions $g_k(t)$ grow as a consequence of parametric instabilities. The fluctuation contribution to the energy density, the last term in eqn. (\ref{enerdensren2}) for $k_m=0$ (no spontaneous symmetry breaking, see eqn. (\ref{enerdensren})) describes the production of adiabatic particles and is positive definite.   Therefore, as a consequence of  conservation of energy the growth
of the fluctuations associated with particle production must result in a drain of energy from the first two terms in (\ref{enerdensren2}), thereby resulting in  damping of the amplitude of $\varphi(t)$. As the amplitude diminishes, the width of the unstable bands diminishes and parametric amplification becomes less efficient but continues until the amplitude vanishes, this is the case for small oscillations as shown by the analysis of Mathieu's equation. Hence we \emph{conjecture} that   this behavior leads to an asymptotic fixed point of equations (\ref{eomfiren2}-\ref{ModeTimeEvoren}) with $\ddot{\varphi}=0;\dot{\varphi}=0$. As the amplitude $\varphi(t)$ diminishes, the analysis based on Mathieu's equation becomes more reliable. As the width of the unstable bands diminish as
a consequence of a diminishing amplitude,  the mode functions approach linear combinations of adiabatic mode functions and the Bogoliubov coefficients (\ref{tilA},\ref{tilB}) become slowly varying functions of time asymptotically becoming constants. In this asymptotic long time limit $\omega_k(\varphi(t)) \rightarrow \omega_k(\infty) = \sqrt{k^2+m^2_R}$ (for the tree level potential (\ref{treeV})) and it follows from equations (\ref{gexpa},\ref{dergexpa}) that
\be {|\dot{g}_k(t)|^2+\omega^2(t)|g_k(t)|^2}_{~\overrightarrow{t\rightarrow \infty}} ~~   {\omega_k(\infty)} \,\Big[ 1+2 \widetilde{\mathcal{N}}_k (\infty)\Big] \,,\label{asycombo}\ee
where we have used equations (\ref{Wroab},\ref{Ngrel}). This \emph{assumption} leads to the following asymptotic form of the energy density (\ref{enerdensren2}) (setting $\hbar=1$),
 \be \mathcal{E}=  V_{eff}(\varphi(\infty))+ \int \frac{d^3k}{(2\pi)^3}\,\omega_k(\infty) \, \widetilde{\mathcal{N}}_k (\infty)  \,. \label{enerdensasy}\ee The occupation numbers $\widetilde{\mathcal{N}}_k(\infty)$ are large for the range of $k$ corresponding to the unstable bands.

 This result is expected as a corollary of the main conjecture: dissipative damping from particle production results in the relaxation of the mean field towards stationary value   $\varphi(\infty)$.  Furthermore, in the asymptotic long time limit
 \be |g_k(t)|^2_{~\overrightarrow{t\rightarrow \infty}} ~~ \frac{1}{2\omega_k(\infty)}\Big[1+2 \widetilde{\mathcal{N}}_k(\infty)\Big] \,,\label{gisasy}\ee where rapidly oscillating terms $\propto e^{\pm 2i\omega_k(\infty)t}$ average out by dephasing and have been neglected.

 The asymptotic value $\varphi(\infty)$   is the solution of the equation of motion with $\ddot{\varphi}=\dot{\varphi}=0$, namely

 \be   \frac{d}{d\varphi}  {V}^R_{eff}(\varphi(\infty);\mu)  \, + \,
  V^{'''}_R(\varphi(\infty))\,\int  \frac{d^3k}{(2\pi)^3} \,\frac{\widetilde{\mathcal{N}}_k(\infty)}{2\omega_k(\infty)} =0 \,.\label{nobs} \ee

 In the case without symmetry breaking, there is the obvious solution $\varphi(\infty)=0$.  The relaxation of the mean field leads to an asymptotic  stationary   state,  with  all the energy of the non-equilibrium initial state transferred to  a highly excited state described by a distribution function $\widetilde{\mathcal{N}}_k (\infty)$. This distribution function  is large in $k-space$ within the unstable resonant bands where adiabatic particles are produced via parametric amplification with larger amplitudes and  bandwidths for smaller $k$. Notice that the asymptotic state must truly be stationary, any small amplitude oscillation will result in parametric amplification and particle production with the concomitant damping of the mean field.

 \vspace{1mm}

 \textbf{ii:) With symmetry breaking:}  Many of the features of the dynamical evolution described above also apply in the case where the (effective) potential allows for symmetry breaking minima away from $\varphi=0$, with the addition of spinodal instabilities and the concomitant particle production.

 Let us consider first the case wherein the initial values of the mean field $\dot{\varphi}(0);\varphi(0)$ lead to oscillations around one of the broken symmetry minima, possibly with excursions into the spinodal region but not over the hump of the potential at its maximum. As the mean field samples the spinodal region in its evolution, the spinodal instabilities lead to the growth of the modes $g_k(t)$ with $k< K_s$ thus draining energy from the first two terms in eqn. (\ref{enerdensren2}) and damping the amplitude of $\varphi(t)$. As the amplitude diminishes, the oscillations no longer probe the spinodal region but while the mean field oscillates around the broken symmetry minimum, there are still parametric instabilities that lead to the growth of $g_k(t)$. Particle production from these instabilities  will continue until the $\varphi(t)$ stops oscillating at the stable minimum at $\varphi(\infty)$, with $\ddot{\varphi}(\infty)=0;\dot{\varphi}(\infty)=0$. Because the minima are stable it follows that $\mathcal{M}^2(\varphi(\infty))>0$, and the oscillation frequencies around these minima   $\omega_k(\infty) = \sqrt{k^2+\mathcal{M}^2(\varphi(\infty))}$ are real. In the asymptotic long time limit
  \be {|\dot{g}_k(t)|^2+\omega^2(t)|g_k(t)|^2}_{~\overrightarrow{t\rightarrow \infty}} ~~   {\omega_k(\infty)} \,\Big[ 1+2 \widetilde{\mathcal{N}}_k (\infty)\Big]\,,  \ee  therefore
 \be {\mathcal{E}_{fR}(t)}_{~\overrightarrow{t\rightarrow \infty}} ~~ \int \frac{d^3k}{(2\pi)^3}\,\omega_k(\infty) \, \widetilde{\mathcal{N}}_k (\infty)+ \int^{k_m}_0 k^2\,\omega_k(\infty)\,\frac{dk}{4\pi^2}\,, \label{bsen}  \ee the last term cancels \emph{exactly} the contribution from the function $\mathcal{F}$ in eqn. (\ref{vefina}), yielding

 \be \mathcal{E} = V_{eff}(\varphi(\infty))+ \int \frac{d^3k}{(2\pi)^3}\,\omega_k(\infty) \, \widetilde{\mathcal{N}}_k (\infty)\,. \label{enebsasy}\ee

 In this case the asymptotic adiabatic particle number $\widetilde{\mathcal{N}}_k(\infty)$ will also have a large population within the spinodally unstable band $k<K_s$, along with
 the parametric amplified bands.

   In the long time limit, the relation (\ref{gisasy}) holds, where contributions from fast oscillating terms average out, and the term $1/2\omega_k(\infty)$ in (\ref{gisasy}) when input into eqn. (\ref{eomfiren2}) cancels the contribution from the function $\mathcal{F}$ to $d\overline{V}^R_{eff}/d\varphi$ yielding the asymptotic solution form of the equation of motion (\ref{eomfiren2})
 \be   \frac{d}{d\varphi}  {V}^R_{eff}(\varphi(\infty);\mu)  \, + \,
  V^{'''}_R(\varphi(\infty))\,\int  \frac{d^3k}{(2\pi)^3} \,\frac{\widetilde{\mathcal{N}}_k(\infty)}{2\omega_k(\infty)} =0 \,,\label{asyeomfiren2} \ee which coincides with (\ref{nobs}) for the case without symmetry breaking.  However, in the case with symmetry breaking, $\varphi(\infty)=0$ is \emph{not} a self-consistent solution because $V^{''}_R(0)<0$ and the mode functions would grow exponentially preventing a stationary solution, which is possible only when $V''(\varphi(\infty)) >0$.   Equation (\ref{asyeomfiren2})
  clearly displays one of the main results: the asymptotic equilibrium value $\varphi(\infty)$ is \emph{not} a minimum of the effective potential, but includes a substantial contribution from particle production.

A similar analysis holds in the case of large initial amplitude $\varphi(0)$. Consider an initial condition wherein the mean field is released from high up in the potential  allowing it to roll down the hill and up   through the spinodal, over the hump at the maximum and over to the other side, rolling down through the spinodal on the other side and up again the potential. Every excursion of the mean field through the spinodal results in a burst of particle production from spinodal instabilities thereby draining energy from the mean field, which eventually will undergo small oscillations around either one of the minima. During the oscillation around the minima  parametric amplification also leads to particle production until the mean field settles at this  minimum  with $\dot{\varphi}=\ddot{\varphi}=0$ and the $g_k(t)$ bound in time. The asymptotic solutions (\ref{enebsasy},\ref{asyeomfiren2}) also describe this case with large initial amplitudes sampling the broken symmetry minima during the evolution until settling down in one of them. The only difference with the small(er) amplitude case described above is in the total energy density and the asymptotic value of $\widetilde{\mathcal{N}}_k(\infty)$ which reflects the different energy densities.

   This analysis leads us to  suggest  a new kind  of \emph{phase diagram}: the asymptotic equilibrium order parameter $\varphi(\infty)$ versus energy density as a characterization of the broken symmetry phases with high energy density.

  The results (\ref{enebsasy},\ref{asyeomfiren2}) taken together have a simple and clear physical interpretation: in absence of particle production $\widetilde{\mathcal{N}}_k(\infty) = 0 \,\, \forall k$, the equilibrium states correspond to
  \be  \frac{d}{d\varphi}  {V}^R_{eff}(\varphi(\infty);\mu) =0 ~~;~~ \mathcal{E} = V_{eff}(\varphi(\infty))\,,\label{asynopp} \ee namely the minimum of the effective potential which includes radiative and renormalization corrections, in fact this was the rationale for the \emph{static} effective potential in the first place. However, under the constraint of \emph{conserved energy density}, the actual asymptotic state must account for the energy transfer from the mean field that has relaxed to equilibrium, to excited states (fluctuations) which are described by the adiabatic particle numbers $\widetilde{\mathcal{N}}_k(\infty) \neq 0$. The asymptotic expectation value is no longer the minimum of the effective potential but is modified by particle production, which in turn depends on the energy density.

  Of course the conjectures on the asymptotic dynamics and emerging stationary states   must be confirmed by a thorough numerical analysis, which is clearly beyond the scope of this article.

 \subsection{Asymptotic excited states: highly entangled two-mode squeezed states.}\label{subsec:asystates}

 As argued above, the asymptotic stationary state is characterized by a distribution function of produced adiabatic particles, $\widetilde{\mathcal{N}}_k(\infty)$. As the evolution of the mean field and quantum fluctuations is described by an initial value problem, we can consider the initial state, determined by the initial conditions (\ref{mfini},\ref{inigs1},\ref{gkaspi},\ref{dotgkaspi}) as the ``in'' state with vanishing occupation number, and the asymptotic stationary state as the ``out'' state. In the transition from the ``in'' to the ``out'' state    the mean field  relaxes to a minimum of the effective potential and the energy density, originally stored in the mean field,  is transferred to excited states (fluctuations), in the form of particle production. At long time, as the mean field relaxes to the asymptotic  equilibrium value $\varphi(\infty)$ solution of the equation (\ref{asyeomfiren2}) (similar to (\ref{nobs})), the oscillation frequencies     are real and   evolve in time slowly as the amplitude of the mean field relaxes to equilibrium, therefore the zero order adiabatic definition of ``particles'' described by equations (\ref{tilA}-\ref{adianum}) reliably describes particles in  the ``out'' state, as discussed in section (\ref{subsec:energy}).

  The Bogoliubov transformation (\ref{cops}) is implemented by a unitary transformation, which is obtained as follows. First write
 \bea \widetilde{A}_k(t) & = &  \cosh(\vartheta_k(t))\,e^{\frac{i}{2}(\theta^+_k(t)+\theta^-_k(t))}~~;~~\tilde{B}_k(t) = \sinh(\vartheta_k(t))\,e^{\frac{i}{2}(\theta^+_k(t)-\theta^-_k(t))}\,\label{ABreds}\\ \tilde{a}_k  & = &  a_k \, e^{\frac{i}{2} \theta^-_k(t)}~~;~~ \tilde{a}^\dagger_{-k}= {a}^\dagger_{-k}\,e^{-\frac{i}{2} \theta^-_k(t)}\,\label{tilas}\\
 \tilde{c}_k(t)  & = &  c_k(t)\, e^{-\frac{i}{2} \theta^+_k(t)}~~;~~ \tilde{c}^\dagger_{-k}(t)= {c}^\dagger_{-k}(t)\,e^{\frac{i}{2} \theta^+_k(t)}\,\label{cilas} \eea where we have used that $\widetilde{A}_k(t);\widetilde{B}_k(t)$ are functions solely of $k^2$. In terms of these definitions and canonically transformed operators, the Bogoliubov transformation (\ref{cops}) becomes
 \be \tilde{c}_{\vk}(t) = \tilde{a}_{\vk}\,\cosh(\vartheta_k(t))+\tilde{a}^\dagger_{-\vk}\,\sinh(\vartheta_k(t))\,.\label{bogos}\ee This transformation is implemented by the following unitary operator
\be S[\vartheta(t)] = \Pi_{\vk}\,\exp\Big\{\vartheta_k(t) \,\Big[\widetilde{a}_{-\vk}\,\widetilde{a}_{\vk}- \widetilde{a}^\dagger_{\vk}\,\widetilde{a}^\dagger_{-\vk} \Big]\Big\}~~;~~ S^{-1}[\vartheta(t)] = S^\dagger[\vartheta(t)]=S[-\vartheta(t)] \,, \label{Strfo}\ee yielding
\be S[\vartheta(t)]\, \widetilde{a}_{\vk}\,S^{-1} [\vartheta(t)]  =    \widetilde{c}_{\vk}(t)\,, \label{bogoc} \ee which can be confirmed by expanding the exponentials, using the identity
 \be
e^{X}Ye^{-X}=Y+[X,Y]+\frac{1}{2!}[X,[X,Y]]+\cdots
\ee
 and the canonical commutation relations.

  An important identity yields the following factorization of the exponential\cite{barnett},
 \bea S[\vartheta] & = &  \Pi_{\vk}\,\exp\Big\{-\ln(\cosh(\vartheta_k)) \Big\}~  \exp\Big\{-\tanh(\vartheta_k)\,\widetilde{a}^\dagger_{\vk}\,\widetilde{a}^\dagger_{-\vk}   \Big\}~ \exp\Big\{-2\ln(\cosh(\vartheta_k))\, \widetilde{a}^\dagger_{\vk}\,\widetilde{a}_{\vk}   \Big\}\nonumber \\
 & \times &  \exp\Big\{\tanh(\vartheta_k)\,\widetilde{a}_{-\vk}\,\widetilde{a}_{\vk}   \Big\} \label{factorS}  \,,  \eea where $\vartheta_k\equiv \vartheta_k(t)$.

 The inverse Bogoliubov transformation is given by
 \bea  \widetilde{a}_{\vk} & = &  \widetilde{c}_{\vk}\,\cosh(\vartheta_k) - \widetilde{c}^\dagger_{-\vk}\,\sinh(\vartheta_k) \nonumber \\
 \widetilde{a}^\dagger_{-\vk} & = & \widetilde{c}^\dagger_{-\vk}\,\cosh(\vartheta_k)- \widetilde{c}_{\vk}\,\sinh(\vartheta_k)\,.  \label{invbogo}\eea The unitary operator that implements it
 is
 \be T[\vartheta] = \Pi_{\vk}\,\exp\Big\{-\vartheta_k \,\Big[\widetilde{c}_{\vk}\,\widetilde{c}_{-\vk} - \widetilde{c}^\dagger_{-\vk}\,\widetilde{c}^\dagger_{\vk} \Big]\Big\}~~;~~ T^{-1}[\vartheta] = T[-\vartheta] \,, \label{Ttrfo}\ee so that

 \bea T[\vartheta]\, \widetilde{c}_{\vk}\,T^{-1} [\vartheta] & = & \widetilde{a}_{\vk} \nonumber \\
 T[\vartheta]\, \widetilde{c}^\dagger_{-\vk}\,T^{-1} [\vartheta] & = & \widetilde{a}^\dagger_{-\vk}\,. \label{Tbog}\eea
 The factorized form of $T[\vartheta]$ is
 \bea T[\vartheta] & = &  \Pi_{\vk}\,\exp\Big\{-\ln(\cosh(\vartheta_k)) \Big\}~  \exp\Big\{\tanh(\vartheta_k)\,\widetilde{c}^\dagger_{\vk}\,\widetilde{c}^\dagger_{-\vk}   \Big\}~ \exp\Big\{-2\ln(\cosh(\vartheta_k))\, \widetilde{c}^\dagger_{\vk}\,\widetilde{c}_{\vk}  \Big\}\nonumber \\
 & \times &  \exp\Big\{-\tanh(\vartheta_k)\,\widetilde{c}_{-\vk}\,\widetilde{c}_{\vk}   \Big\} \label{factorT}  \,.  \eea  With the instantaneous (zeroth order) adiabatic vacuum state $\ket{0_a(t)}$  defined such that
 \be c_k(t) \ket{0_a(t)} = 0 ~~ \forall k, t \,.\label{advac}\ee

  The  operator $T[\vartheta]$  allow us to relate the ``adiabatic'' vacuum state $\ket{0_a(t)}$  to the coherent state $\ket{\Phi}$ (annihilated by $a_k$).   Pre-multiplying  (\ref{advac}) by $T[\theta]$ and inserting $T^{-1}[\theta]\,T[\theta]=1$,  yields
 \be \underbrace{\Big( T[\vartheta]\,c_{\vk}\,T^{-1}[\theta]\Big)}_{a_{\vk}}\, \Big(T[\vartheta]\,\ket{0_a(t)}\Big)  =0  \,,  \label{Toutvac}\ee from
 which   the relation between vacua follows, namely
 \be \ket{\Phi} = T[\vartheta]\,\ket{0_a(t)}\,.  \label{relavacua}\ee
 Therefore, we find
\be \ket{\Phi} = \Pi_{\vk}\Bigg\{\Big[\cosh(\vartheta_k)\Big]^{-1}~ \,\sum_{n_{\vk}=0}^\infty \Bigg(e^{i\theta^+_k}\,\tanh(\vartheta_k) \Bigg)^{n_{\vk}} \ket{n_{\vk}\,; \, {n}_{-\vk}}\Bigg\} \,, \label{vacin}\ee where the adiabatic  particle-pair states
\be  \ket{n_{\vk}; {n}_{-\vk}}  = \frac{\Big(c^{\dagger}_{\vk}\Big)^{n_{\vk}}}{\sqrt{n_{\vk}!}}~
\frac{\Big(c^{\dagger}_{-\vk}\Big)^{n_{\vk}}}{\sqrt{n_{\vk} !}} \ket{0_a} ~~;~~ n_{\vk} = 0,1,2 \cdots\,. \label{noutsbose}\ee

In quantum optics these correlated states are known as two-mode squeezed states\cite{barnett}, where as discussed in section (\ref{subsec:energy}) the   Fock states
\be  \ket{n_{\vk}(t)}  = \frac{\Big(c^{\dagger}_{\vk}(t)\Big)^{n_{\vk}}}{\sqrt{n_{\vk}!}}~  \ket{0_a(t)}\,,\label{fsa}\ee are instantaneous eigenstates of the Hamiltonian (\ref{adhdel}) with eigenvalue $\hbar \omega_k(t)(n_k(t)+1/2)$.

We note that the Fock pair states (\ref{noutsbose}) are eigenstates of the \emph{pair number operator}
\be \widehat{\eta}_{\vk} = \sum_{m_{\vk}=0}^{\infty} m_{\vk}\,\, \ket{m_{\vk}; {m}_{-\vk}}\bra{m_{\vk}; {m}_{-\vk}}  \,, \label{pairop} \ee namely
\be \widehat{\eta}_{\vk} \, \ket{n_{\vk}; {n}_{-\vk}} = n_{\vk} \,\ket{n_{\vk}; {n}_{-\vk}}~~;~~ n_{\vk} = 0,1,2\cdots \,.  \label{eigenNk}\ee

 Several checks are in order:
\be \langle \Phi|\Phi\rangle = \Pi_{\vk} \frac{1}{\cosh^2(\vartheta_k)}\, \sum^{\infty}_{n_k=0}(\tanh^2(\vartheta_k))^{n_k} = \Pi_{\vk}  \frac{1}{\cosh^2(\vartheta_k)}\,\frac{1}{1-\tanh^2(\vartheta_k)} = 1 \,, \label{ck1}\ee

\be \bra{\Phi}c^\dagger_{\vp}c_{\vp}\ket{\Phi} =      \frac{1}{\cosh^2(\vartheta_p)}\, \sum^{\infty}_{n_p=0}n_p\, (\tanh^2(\vartheta_p))^{n_p} = \sinh^2(\vartheta_p) = |\widetilde{B}_p|^2 =\widetilde{\mathcal{N}}_p\,, \label{ck2}\ee

Therefore, in terms of the asymptotic adiabatic ``out'' particle states, the coherent state $\ket{\Phi}$ is a strongly correlated, entangled state of back-to-back pairs of particles with occupation numbers $\widetilde{\mathcal{N}}_k$ populated in bands: for $k\leq K_s$ for spinodally produced particles and the unstable bands for the particles produced by parametric amplification.

\subsection{Decoherence and entropy:}\label{subsec:decoherence}

For large energy density  the occupation numbers in the bands of instability are expected to be large with a continuum distribution in each band as the energy is transferred from the mean field to the excitations described by the adiabatic particle states. This transfer of energy from a single mode, the mean field,  to a continuum  of states in the various bands, each with finite bandwidth in momentum,  \emph{intuitively suggests} the emergence of entropy.

However, the density matrix
\be \hat{\rho}= \ket{\Phi}\bra{\Phi}\,,\label{densmtx}\ee describes a pure state and is time independent in the Heisenberg picture. In the basis of the asymptotic ``out'' adiabatic particle states, it is given by
\be \hat{\rho} = \Pi_{\vk} \Pi_{\vp}  \sum_{n_{\vk}=0}^\infty  \sum_{m_{\vp}=0}^\infty \mathcal{C}^*_{m_{\vp}}(\vp)~ \mathcal{C}_{n_{\vk}}(\vk)~\ket{n_{\vk}; {n}_{-\vk}}\bra{m_{\vp}; {m}_{-\vp}} \,,\label{rhosout}\ee where
\be \mathcal{C}_{n_{\vk}}(\vk) = \frac{\Bigg(e^{i\theta^+_k}\,\tanh(\vartheta_k) \Bigg)^{n_{\vk}}}{\cosh(\vartheta_k)}\,, \ee and the angles $\theta^+_k;\vartheta_k$ correspond  to the
asymptotic values with $\varphi(\infty)$.

The diagonal elements of the density matrix   are given by the probabilities of finding a back-to-back pair of $n_{\vk}$ adiabatic particles, namely
\be P_{n_{\vk}} = | \mathcal{C}_{n_{\vk}}(\vk)|^2 = \frac{\big(\widetilde{\mathcal{N}_k}(\infty)\big)^{\,n_{\vk}}}{\big({1+\widetilde{\mathcal{N}_k}(\infty)}\big)^{\,1+n_{\vk}}} \,,\label{Pns}\ee remarkably,
this form of the diagonal matrix elements is similar to that of a thermal density matrix in the basis of (free) Fock quanta, but with $\widetilde{\mathcal{N}}_{\vk}(\infty)$ replaced by the Bose Einstein distribution function.

Consider a Heisenberg picture operator $\mathcal{O}_{\delta}(t)$  associated with an observable related to the fluctuation operator $\hat{\delta}$, which by dint of the expansion (\ref{adiaexp}) at long time is associated with the asymptotic ``out'' adiabatic particle states. Asymptotically when the mean field has relaxed to its equilibrium value $\varphi(\infty)$ the  Hamiltonian $H_\delta(t)$ given by (\ref{adhdel}) becomes time independent, therefore the time evolution of the Heisenberg picture operator $\mathcal{O}_{\delta}(t)$ is given by
\be \mathcal{O}_{\delta}(t) = e^{iH_\delta (t-t_0)}\,\mathcal{O}_{\delta}(t_0)\, e^{-iH_\delta (t-t_0)}\,,\label{Oasy}\ee where $t_0$ is a late time at which the mean field has relaxed to equilibrium, and $t \gg t_0$. The expectation value of $\mathcal{O}_\delta$ in the density matrix (\ref{densmtx}) is given by
\be \bra{\Phi}\mathcal{O}_{\delta}(t)\ket{\Phi} = \mathrm{Tr}\mathcal{O}_{\delta}(t_0)\hat{\rho}(t) \,,\label{expO} \ee where the time dependent density matrix in the Schroedinger picture is given by

 \be \hat{\rho}(t) =  e^{-iH_\delta (t-t_0)}\hat{\rho}(t_0)  e^{iH_\delta (t-t_0)}~~;~~ \hat{\rho}(t_0)=\ket{\Phi}\bra{\Phi} \,.\label{rhooft}\ee Since the zeroth order adiabatic ``out'' states are (instantaneous) eigenstates of $H_\delta$ it follows that
 \be \hat{\rho}(t)  = \Pi_{\vk} \Pi_{\vp}  \sum_{n_{\vk}=0}^\infty  \sum_{m_{\vp}=0}^\infty \mathcal{C}^*_{m_{\vp}}(\vp)~ \mathcal{C}_{n_{\vk}}(\vk)~\ket{n_{\vk}; {n}_{-\vk}}\bra{m_{\vp}; {m}_{-\vp}}\,e^{-iW_{n,m}(t-t_0)}\,,\label{rhosoutoft}\ee where
 \be W_{n,m}= 2\,\Big(n_k\,\omega_k(\infty)-m_p\,\omega_p(\infty)\Big)  \,.\label{frecs}\ee The off-diagonal matrix elements in the adiabatic ``out'' basis are a manifestation of coherence, and unitary time evolution.

  At long time $t \gg t_0$, the off diagonal terms with $n_k \neq m_p; k\neq p$ oscillate very rapidly, the continuum of modes within each band fall out of phase  leading to rapid dephasing and averaging out. In fact taking a long time average of the expectation value (\ref{expO}),
 \be   \frac{1}{T} \int^T_{t_0}  \mathrm{Tr}\mathcal{O}_{\delta}(t_0)\hat{\rho}(t) \, dt_{~~\overrightarrow{T\rightarrow \infty}~~}\mathrm{Tr}\mathcal{O}_{\delta}(t_0)\hat{\rho}^{(d)} \,,  \ee where $\hat{\rho}^{(d)}$ is diagonal in the Fock ``out'' basis of correlated --entangled-- pairs, namely
 \be \hat{\rho}^{(d)} = \Pi_{\vk}       \sum_{n_{\vk}=0}^\infty   P_{n_{\vk}}~\ket{n_{\vk}; {n}_{-\vk}}\bra{n_{\vk}; {n}_{-\vk}}\,,\label{rhodiag}\ee with the probabilities
 (\ref{Pns}). The diagonal density matrix $\hat{\rho}^{(d)}$  describes a \emph{mixed state}. The main ingredient in this analysis is that the ``out'' adiabatic particle states are (instantaneous) eigenstates of $H_\delta$ and that each band has a continuum of modes each evolving in time with different frequency, leading to dephasing and decoherence in the long time limit.

 This argument, based on \emph{decoherence by dephasing} at long time yielding a density matrix diagonal in the ``energy'' basis underpins the  \emph{eigenstate thermalization hypothesis}\cite{sred,rigol,deutsch} and is at the heart of the arguments on thermalization in closed quantum systems, a subject of much current theoretical and experimental interest.

  The entropy associated with this mixed state can be calculated simply by establishing contact between the density matrix $\rho^{(d)}$ and that of  quantum statistical mechanics in equilibrium described by a fiducial Hamiltonian
\be \widehat{\mathcal{H}} = \sum_{\vk}  {E}_k\, \widehat{\eta}_{\vk} \,,\label{fiduH} \ee with $\widehat{\eta}_{\vk}$ the pair number operator (\ref{pairop}) with eigenvalues $n_{\vk} = 0,1,2\cdots$,   and the fiducial (dimensionless) ``energy''
\be  {E}_k = - \ln\big[\tanh^2(\vartheta_k) \big]\,, \label{fiduenergy}\ee which suggestively yields the distribution function
\be \widetilde{\mathcal{N}}_{\vk}(\infty) = \frac{1}{e^{ {E}_k}-1}\,.\label{ther}\ee

This fiducial Hamiltonian (\ref{fiduH}) is diagonal in the correlated basis of particle-antiparticle pairs, it should not be confused with the Hamiltonian $H_\delta$ of eqn. (\ref{adhdel}), they act on different Hilbert spaces and feature different eigenvalues. The main purpose of the fiducial Hamiltonian $\widehat{\mathcal{H}}$ is to  identify
\be \hat{\rho}^{(d)}  = \frac{e^{-\widehat{\mathcal{H}}}}{\mathcal{Z}}~~;~~ \mathcal{Z}= \mathrm{Tr}\,e^{-\widehat{\mathcal{H}}} \equiv e^{-\mathbb{F}} \,,\label{rhofidu}\ee with $\mathbb{F}$ the fiducial (dimensionless) free energy, and the partition function
 \be \mathcal{Z} =  \Pi_{\vk} \mathcal{Z}_{\vk}~~;~~ \mathcal{Z}_{\vk} = \frac{1}{\Big[ 1- e^{-{E}_{k}}\Big]} = \frac{1}{\Big[ 1- \tanh^2(\vartheta_k)\Big]} \,,   \label{fiduzexa}\ee thereby establishing a direct relation to a problem in quantum statistical mechanics.

  Since $\widehat{\mathcal{H}}$ is diagonal in the basis of the pair Fock states, so is $\hat{\rho}^{(d)}$, and obviously the matrix elements of (\ref{rhofidu})  in the pair basis are identical to those of (\ref{rhodiag}), with the identification of the pair probability (\ref{Pns}) as

\be P_{n_{\vk}}   = \frac{e^{- {E}_k\,n_{\vk}}}{\mathcal{Z}_{\vk}}= \frac{\big(\widetilde{\mathcal{N}_k}(\infty)\big)^{\,n_{\vk}}}{\big({1+\widetilde{\mathcal{N}_k}(\infty)}\big)^{\,1+n_{\vk}}}  \,. \label{prob2} \ee

 The von Neumann entropy associated with this mixed state is
 \be S  = - \mathrm{Tr}\,\rho^{(d)} \,\ln \rho^{(d)}  \,. \label{vNSd} \ee

 The eigenvalues of $\rho^{(d)} $ are the probability for each state of $n_{\vk}$ pairs of momenta $(\vk;-\vk)$, namely $P_{n_{\vk}}$    therefore  the von Neumann entropy is given by
\be S  = -\sum_{\vk}\sum_{n_{\vk}=0}^\infty P_{n_{\vk}}\,\ln P_{n_{\vk}}\,. \label{Sden} \ee A straightforward calculation yields the entropy \emph{density}\footnote{The entropy can also be calculated with the analogy $\mathbb{F} = U-S $  , with $U=\mathrm{Tr}\mathcal{H}\hat{\rho}^{(d)}$ as in statistical mechanics.}
\be s = \int \Big[\big(1+ \widetilde{\mathcal{N}}_{\vk}(\infty)\big)\,\ln\big(1+ \widetilde{\mathcal{N}}_{\vk}(\infty)\big)- \widetilde{\mathcal{N}}_{\vk}(\infty))\ln  \widetilde{\mathcal{N}}_{\vk}(\infty)  \Big] \, \frac{d^3k}{(2\pi)^3}\,.\label{entrodens}\ee Remarkably the entropy features the same form as in a quantum free thermal Bose gas but with the equilibrium distribution functions replaced by the asymptotic distribution functions of the produced ``out'' adiabatic particles.

Although the similarity with quantum statistical mechanics in thermal equilibrium is striking, we emphasize that the distribution functions are non-thermal and localized in bands in momentum.

This entropy is a direct corollary of the conjecture on the emergence of an asymptotic stationary state with a large population of
adiabatic ``out'' particles. These are the eigenstates of the evolution Hamiltonian for the fluctuations,   which asymptotically becomes time independent.
Decoherence by dephasing in the basis of energy eigenstates is one of the main arguments towards the description of microcanonical quantum statistical mechanics, and as mentioned above the cornerstone of the eigenstate thermalization hypothesis, that describes thermalization in closed quantum systems.

The diagonal form of the density matrix (\ref{rhodiag})   also emerges from tracing over one member of the correlated  pair states in the full density matrix (\ref{rhosoutoft}), therefore \emph{formally} the entropy (\ref{Sden}) is equivalent to the entanglement entropy. Although in the cases studied above we focused on neutral scalar fields, if instead the fields feature a charge quantum number, and the pair states are of particle and antiparticle, tracing over either of them would yield an entanglement entropy similar to (\ref{Sden}).

\vspace{1mm}

\section{Conclusion and further questions:}\label{conclusion}

The effective potential is a very useful concept to understand the \emph{equilibrium} phase structure of a theory, in particular spontaneous symmetry breaking, including
quantum and thermal corrections. Although it is defined to describe static phenomena, it is often used to study the dynamical evolution of the expectation value
of a field. Motivated by its ubiquitous use in phenomenological approaches to dynamical evolution, including in cosmology, our objectives  in this article are to  critically examine  whether using the effective potential to study the dynamics of a coherent mean field, or expectation value is warranted, and  to  provide a consistent  framework to study its evolution when it is not. We implemented a Hamiltonian formulation to obtain the energy functional up to one loop which yields the static effective potential and extended it to obtain the equation of motion for the expectation value of a scalar field in the dynamical case. This formulation is manifestly energy conserving and renormalizable. We introduced an adiabatic approximation to establish if a quasi-static evolution warrants the use of the static effective potential in the equations of motion and found that doing so implies an explicit violation of energy conservation. Furthermore the regime of validity of such an adiabatic approximation is severely restricted. Breakdown of adiabaticity is recognized in two ubiquitous instances of fundamental and phenomenological relevance: parametric amplification associated with instabilities from resonant excitations by oscillating mean fields and spinodal decomposition, instabilities stemming from  the growth of correlations during phase transitions in the case of spontaneous symmetry breaking.

The breakdown of adiabaticity is directly linked to the production of adiabatic particles, which we show to describe the asymptotic ``out'' state at long time. A self-consistent, energy conserving and renormalizable framework that is amenable to numerical implementation is introduced. Energy conservation implies the emergence of asymptotic stationary states described by highly excited entangled adiabatic particle states. Their  distribution functions   are localized in momentum space in regions of spinodal or parametric instabilities.
In the case when the tree level potential admits broken symmetry minima, the asymptotic value of the order parameter is \emph{not} the minima of the effective potential, but receives corrections from the excited states, and the energy density transferred to these via particle production.  This led us to conjecture on the characterization of phases in terms of novel ``phase diagrams'' of \emph{asymptotic expectation values of the scalar field, namely the order parameter, versus energy density}.

Although we considered simple examples of tree level potentials to anchor the discussions, the results are of far broader significance. Parametric and spinodal instabilities are ubiquitous in theories without and with symmetry breaking,  and generally call into question the applicability of the effective potential to study the dynamics of coherent mean fields.

The asymptotic stationary states are fixed points of the dynamics corresponding to equilibria compatible with the constraint of fixed energy (energy conservation). These novel equilibria are non-universal as they depend on couplings, parameters and initial conditions on $\varphi,\dot{\varphi}$ and mode functions that determine the energy density.  In the case of tree level potentials featuring broken symmetry minima, the asymptotic equilibrium values of the mean field are very different from that obtained from the effective potential, a consequence of profuse particle production.  The distribution functions of adiabatic particles are non-thermal and non-universal, peaked at bands corresponding to spinodally and or parametrically produced particles, since at this level (one loop) of approximation there are no collision terms that would redistribute energy and momenta away from the instability bands. A direct corollary of the emergence of an asymptotic state, is decoherence by dephasing of the Schroedinger picture density matrix in the basis of the asymptotic ``out'' adiabatic particle states, and the concomitant emergence of entropy, surprisingly the form of the entropy is similar to that of a free quantum Bose gas but in terms of the distribution function of the produced particles.

Our study has been restricted to the one-loop approximation to  compare with the familiar one-loop effective potential and exhibit its shortcomings to describe the dynamics in the simplest and clearest example.  Our main results are of broader significance and transcend the particular approximation: \textbf{i:)} the effective potential is ill suited to study dynamics, \textbf{ii:)}  there is a substantial transfer of energy of the mean field to excitations, these are described in terms of asymptotic ``out'' states based on the zeroth adiabatic modes, \textbf{iii:)} an asymptotic stationary state must emerge at long time as a consequence of energy conserving dynamics when parametric and or spinodal instabilities occur, \textbf{iv:)} the asymptotic equilibrium value of the mean field is \emph{not} described correctly by the effective potential but also receives corrections from the excited states. This is an unambiguous consequence of energy conserving dynamics, \textbf{v:)} a corollary of  the   asymptotic stationary state is that there emerges an entropy from decoherence and dephasing of the Schroedinger picture density matrix. These are all results that do not depend on the level of approximation, but stem fundamentally from energy conserving dynamics associated with particle production from the evolution of the mean field.

 These results   justify the study of its extension beyond one loop within a manifestly renormalizable and energy conserving framework both to confirm the main conclusions and also to reveal   quantitative characteristics of the approach to the asymptotic state.  A possible avenue would be to include back reaction self-consistently for example within a Hartree-type approximation\cite{boyadiss,boyaspino} which, however, would not include collisions. An alternative would be to implement the effective action approach advocated in the seminal work of ref.\cite{cornwall}.

Non-equilibrium fixed points (or nearly fixed points of the dynamics) have been identified in previous studies within a different framework\cite{berges} including collisional processes, and more recently the dynamics of condensates have been included in Boltzmann equations\cite{wen}.   These approaches could provide an alternative confirmation of the emergence of an asymptotic stationary state and  of a coarse grained entropy in the asymptotic regime as a consequence of decoherence via dephasing in a closed quantum system with energy conserving and unitary dynamics\cite{serreau}, and can shed light on the question if such entropy becomes the thermal entropy.

While our study has been carried out in Minkowski space time, we expect that the results   also have broad impact in cosmology: in the equations of motion for a scalar (or pseudoscalar field), during the time when the Hubble expansion rate $H$ is much larger than the mass, damping from cosmological expansion \emph{may} justify the use of a static effective potential within this time window. However when $H$ becomes much smaller than the mass, oscillations ensue with the concomitant particle production and parametric amplification. We highlighted that the breakdown of adiabaticity is primarily associated with  long wavelength excitations, hence it is important to assess the contribution from super-Hubble modes to the fluctuation contributions to the equations of motion, even during the time window when Hubble friction dominates. Cosmological particle production arising from the energy transfer from mean fields to fluctuations has important consequences in cosmology, as the full energy momentum tensor would feature two components, a ``cold'' component from the coherent mean field, and a ``hotter'' component from the particles produced from either spinodal or parametric instabilities. This possibility warrants further study of the processes described in this work applied to cosmology and on which we will report in future work. Furthermore, extending the treatment to gauge theories will require a clear understanding of gauge invariance in the dynamics and renormalization aspects, these are also topics beyond the scope of this article and subject of future work.

\acknowledgements
  S.C. and  D. B. gratefully acknowledge  support from the U.S. National Science Foundation through grant   NSF 2111743.

\appendix
\section{Instability bands $\kappa^2_{n,\pm}(\alpha)$ for eqn.(\ref{mathieu}).  } \label{app:bands}
From the results in refs.\cite{mathieu1,abra,kova} we obtain the following  power series expansion in $\alpha$ for the band edges $\kappa^2_{n,\pm}$, valid in the range   $0 \leq \alpha \lesssim 2$, the range of validity may be extended by including higher orders in the expansion\cite{mathieu1,kova}.

\bea \kappa^2_{2,-} & = & 3-2\alpha - \frac{\alpha^2}{12}+\frac{5\alpha^4}{13824}-\frac{289\alpha^6}{79626240}+ \cdots \nonumber \\
\kappa^2_{2,+} & = & 3-2\alpha + \frac{5\alpha^2}{12}-\frac{763\alpha^4}{13824}+\frac{1002401\alpha^6}{79626240}+ \cdots \nonumber \\
   \kappa^2_{3,-} & = & 8-2\alpha + \frac{\alpha^2}{16}-\frac{ \alpha^3}{64}+\frac{13\alpha^4}{20480}+ \frac{ 5\alpha^5}{16384}-\frac{1961 \alpha^6}{23592960}\cdots \nonumber \\
\kappa^2_{3,+} & = & 8-2\alpha + \frac{\alpha^2}{16}+\frac{ \alpha^3}{64}+\frac{13\alpha^4}{20480}- \frac{ 5\alpha^5}{16384}-\frac{1961 \alpha^6}{23592960}\cdots \nonumber \\
  \kappa^2_{4,-} & = & 15-2\alpha +\frac{\alpha^2}{30}-\frac{317 \alpha^4}{864000}+\frac{10049\alpha^6}{2721600000}+ \cdots \nonumber \\
\kappa^2_{4,+} & = &  15-2\alpha +\frac{\alpha^2}{30}+\frac{433 \alpha^4}{864000}-\frac{5701\alpha^6}{2721600000}+ \cdots \label{kbands}
\eea

\end{document}